\documentclass[10pt]{scrartcl}
\usepackage{graphicx}
\usepackage[colorlinks=true,citecolor=blue,linkcolor=black,urlcolor=blue]{hyperref}
\usepackage{amsmath,amssymb,amsthm}
\usepackage{comment}
\usepackage{esint}
\usepackage{fancyhdr}
\usepackage{fullpage}
\usepackage{enumerate}
\usepackage[font={small}]{caption}
\def\DMO{\DeclareMathOperator}

\DMO{\Ai}{Ai}
\DMO{\Be}{Be}
\DMO{\Prob}{Prob}
\DMO{\Var}{Var}
\renewcommand{\P}{\mathbb{P}}
\newcommand{\Z}{\mathbb{Z}}

\newcommand{\R}{\mathbb{R}}

\def\E{\mathbb E}

\newcommand{\tr}{\mathrm{tr}}
\newcommand{\normord}[1]{:\mathrel{\mkern2mu #1 \mkern2mu}:}
\newcommand{\ket}[1]{|\mathbin{#1} \rangle}
\newcommand{\bra}[1]{\langle\mathbin{#1}|}

\def\ind{\mathbf{1}}

\renewcommand{\Re}{\mathrm{Re}}
\newcommand{\dd}{\mathrm{d}}

\theoremstyle{definition}

\numberwithin{equation}{section}

\title{Interface fluctuations associated with \\
split Fermi seas}
\date{\vspace{-5ex}}
\author{
  Harriet Walsh\thanks{\, Université d’Angers, CNRS, LAREMA, SFR MATHSTIC, F-49045 Angers \newline \href{mailto:harriet.walsh@univ-angers.fr}{harriet.walsh@univ-angers.fr} 
  \newline
 This project has received funding from the European Research Council (ERC) under the European Union’s Horizon 2020 research and innovation programme under Grant Agreement No. ERC-2017-STG 759702, ``COMBINEPIC'', and from the Agence Nationale de la Recherche via the grant ANR-18-CE40-0033 ``Dimers'' and the Centre Henri Lebesgue  ANR-11-LABX-0020-01.} }

\begin{document}

\maketitle

\begin{center}
{\Large \today{}}
\end{center}
\bigskip

\begin{abstract}
 We consider the asymptotic behaviour of a family of unidimensional lattice fermion models, which are in exact correspondence with certain probability laws on  partitions and on unitary matrices. These models exhibit limit shapes, and in the case where the bulk of these shapes are described by analytic functions, the fluctuations around their interfaces have been shown to follow a universal Tracy--Widom distribution or its higher-order analogue. Non-differentiable bulk limit shape functions arise when a gap appears in some quantum numbers of the model, in other words when the Fermi sea is split. We show that split Fermi seas give rise to new interface fluctuations, governed by integer powers of universal distributions. This breakdown in universality is analogous to the behaviour of a random Hermitian matrix when the support of its limiting eigenvalue distribution has multiple cuts, with oscillations appearing in the limit of the two-point correlation function. We show that when the Fermi sea is split in the lattice fermion model, there are multiple cuts in the eigenvalue support of the corresponding unitary matrix model.
\end{abstract}

\section{Introduction}

\subsection{Context}

Several models in statistical physics and combinatorics have universal interface behaviour which coincides with the behaviour of the largest eigenvalue of a random Hermitian matrix in the Gaussian unitary matrix ensemble (GUE). A typical situation might be described as follows. As the scale of the model grows large, a deterministic limit shape emerges, demarcating the region where the probability of finding the ``particles'' of the model is non-trivial (see e.g.~\cite{Kenyon_lectures,Stephan_lectures}). Suppose that, on a length scale where the typical distance between particles is of order one, the edge of the limit shape is at a distance $L$. The particle probability density generically vanishes at the edge with an exponent of $\frac{1}{2}$ (coinciding with the edge of Wigner's semi-circle law from the GUE case). Then, the fluctuations in the position of the furthest particle typically scale with $L^{\frac{1}{3}}$ around the edge and are asymptotically governed by the \emph{GUE Tracy--Widom distribution}~\cite{Tracy_Widom_1993}, which can be defined in terms of the maximum eigenvalue $\xi_{\max}$ of an $N\times N$ GUE random matrix as 
\begin{equation}
F_{\text{GUE}}(s) := \lim_{N \to \infty} \P \bigg( \frac{\xi_{\max} - (2N)^{\frac{1}{2}}}{2^{-\frac{1}{2}}N^{-\frac{1}{6}}} < s\bigg).
\end{equation}
This notably describes extreme positions of free fermions in confining traps in 1D~\cite{Eisler_2013,Deleporte_Lambert_2021}, and even extends to some   interacting fermion models~\cite{Stephan_2019}; the same interface fluctuations have been found in 2D random growth models~\cite{Praehofer_Spohn_2002}, including by experiment~\cite{Takeuchi_2011}, and for random integer partitions under the Plancherel measure~\cite{Baik_Deift_Johansson_1999} (where the edge is the first part, which has the same law as the length of the longest increasing subsequence of a uniform random permutation).

Analogues of the Tracy--Widom distributions associated with vanishing exponents of $\frac{1}{2m}$ and fluctuations at a scale of $L^{\pm\frac{1}{2m+1}}$ for any positive integer $m$ have recently been identified, defined by the Fredholm determinants 
\begin{equation} \label{eq:Forderm}
F_{2m+1}(s) := \det (1- \mathcal{A}_{2m+1})_{L^2([s,\infty))} = \sum_{n=0}^\infty \frac{(-1)^n}{n!} \int_{s}^{\infty} \hspace*{-1em}\cdots\hspace*{-0.2em} \int_{s}^{\infty} \det_{1 \leq i,j \leq n}  \mathcal{A}_{2m+1}(x_i,x_j) \dd x_1 \cdots \dd x_n
\end{equation}
where $\mathcal{A}_{2m+1}$ is the {order-$m$ Airy kernel}, given by
\begin{equation}   \label{eq:Airyorderm}
\mathcal{A}_{2m+1}(x,y) = \int_{0}^{\infty} \Ai_{2m+1}(x- v) \Ai_{2m+1}(y - v) \dd v
\end{equation}  
in terms of the order-$m$ Airy function 
\begin{equation} \label{eq:Airyfnm}
\Ai_{2m+1}(x) := \frac{1}{2\pi i} \int_{1 + i\R} \exp \bigg[ (-1)^{m-1}\frac{\zeta^{2m+1}}{2m+1} -x \zeta  \bigg] \dd \zeta. 
\end{equation}
These distributions were first found by Le Doussal, Majumdar and Schehr~\cite{LDMS_2018} for the fluctuations in the largest momentum of free fermions in tuned confining potentials. For $m=1$,~\eqref{eq:Forderm} is the Fredholm determinant expression for $F_{\text{GUE}}$; for $m>1$, which is classified as \emph{multicritical} edge behaviour, the distributions $F_{2m+1}$ have not yet been defined in terms of matrix ensembles. These distributions generalise a connection between $F_{\text{GUE}}$ and the Painlevé II equation: Cafasso, Claeys and Girotti~\cite{Cafasso_Claeys_Girotti_2019} proved that each $F_{2m+1}$ encodes a solution of the $m$th equation of the Painlevé II hierarchy (see also~\cite[Appendix~G]{LDMS_2018}). 

This paper revisits a family of lattice fermion models, or equivalently measures on integer partitions, previously shown by Betea, Bouttier and the author~\cite{FPSAC_preprint_BBW,BBW_2023} to universally exhibit order-$m$ Tracy--Widom fluctuations for some $m$, under the hypothesis that a certain ``Fermi sea'' of the model has a single cut. Independent work by Kimura and Zahabi~\cite{Kimura_Zahabi_2-2021b,Kimura_Zahabi_2021} found the same edge behaviour in models of this kind, using a somewhat different approach. The asymptotic behaviour of these models can be studied exactly, as they are determinantal point processes; in the single cut Fermi sea case, the distributions $F_{2m+1}$ arise from a universal edge scaling limit of the two-point correlation function.  Here, we show that this universality picture breaks for the case where the Fermi sea has several cuts. In this case, we find limiting particle densities with the \emph{same} vanishing exponents of $\frac{1}{2m}$ and fluctuation exponents of $\frac{1}{2m+1}$, but \emph{different} limiting distributions, given by integer powers of the distributions $F_{2m+1}$. 
In this case the two-point correlation function does not have an edge scaling limit. Instead it exhibits oscillations, resembling a phenomenon found for multi-cut Hermitian matrix models in~\cite{Bonnet_David_Eynard_2000}. We also show an exact correspondence between lattice fermion models with multi-cut Fermi seas and multi-cut unitary matrix models.

\subsection{Main model and related models}
We primarily consider a model of spinless free fermions on an infinite unidimensional lattice. The same model is described for instance in~\cite[Section~2.1]{BBW_2023}. Labelling the sites by the half integers $\pm \frac12, \pm \frac32, \pm \frac52, \ldots$ for convenience, we work in the second quantisation formalism and let $c_k^\dagger, c_k$ denote the anticommuting fermionic creation and annihilation operators on the site labelled $k$. Then, we consider the ground state of the Hamiltonian
\begin{equation}
H_{\theta \gamma} := \sum_{k \in \Z + \frac12} \bigg[ k \normord{c_k^\dagger c_k} -  \theta \sum_{r \geq 1}  r \gamma_r (c_k^\dagger c_{k+r} + c_k^\dagger c_{k-r})  \bigg]  + \theta^2 \sum_{r\geq 1}  r^2 \gamma_r^2
\end{equation}
where $\theta$ is a positive parameter,  $\gamma = (\gamma_1,\gamma_2,\ldots)$ is a sequence of real numbers with finite support, and $\normord{\cdot}$ denotes normal ordering with respect to the ``domain wall'' state $\ket{\emptyset}$ in which all negative sites are filled and all positive ones are empty, so that 
\begin{equation}
\normord{c_k^\dagger c_k} := c_k^\dagger c_k - \bra{\emptyset} c_k^\dagger c_k \ket{\emptyset} = \begin{cases}c_k^\dagger c_k, &\quad k>0 \\ - c_k c_k^\dagger,  &\quad k<0\end{cases}. 
\end{equation}
Thanks to the normal ordering, this Hamiltonian has finite eigenvalues in spite of the linear potential in the first term, once we only consider states with finitely many empty negative sites and finitely many filled positive sites. Picturing the lattice sites on a horizontal line (with negative labels to the left and positive ones to the right), this means the states we consider have a rightmost occupied site and a leftmost empty site. The second term of the Hamiltonian introduces translation-invariant hopping dynamics between sites a finite distance apart, weighted by the coefficients $\gamma_r$ (which can be positive or negative). The final scalar term shifts the spectrum conveniently.

This model is particularly useful from a universality perspective as it is in exact or asymptotic correspondence with some other interesting models, in particular:

\paragraph{(i) Hermitian Schur measures:} Suppose that we measure the occupation number of every lattice site in the ground state of $H_{\theta \gamma}$ at once, and obtain a random infinite sequence $S$ of the labels of sites observed to be occupied. Then the number of positive elements of $S$ is equal to the number of negative half-integers not in $S$, and for some integer partition $\lambda = (\lambda_1 > \lambda_2 > \ldots)$ we have $S= \{\lambda_i - i + \tfrac{1}{2}, i \in \Z_{>0}\}$ (in our conventions, $\lambda$ is an infinite sequence of non-negative integers which are eventually zero). The law of $S$ may be explicitly given in terms of a partition  by 
\begin{equation} \label{eq:schurmeasS}
\P_{\theta \gamma}(S = \{\lambda_i - i + \tfrac{1}{2}, i \in \Z_{>0}\}) = e^{-\theta^2 \sum_{r}  r \gamma_r^2} s_{\lambda}[\theta \gamma]^2
\end{equation}
where $s_{\lambda}[\theta \gamma]$ denotes the Schur function indexed by $\lambda$ evaluated at the Miwa times $\theta\gamma_1, \theta \gamma_2, \ldots$ (see Appendix~\ref{app:schur} for definitions). Interpreted as a law on partitions, $\P_{\theta\gamma}(\lambda)$ is an instance of Okounkov's \emph{Schur measures}~\cite{Okounkov_2001}, in the ``Hermitian'' case where the two Schur functions are evaluated on complex conjugate sequences. Let us note an immediate parallel between the partition and fermion pictures: if a random partition $\lambda$ is distributed by $\P_{\theta\gamma}$, its shifted first part $\lambda_1 - \frac{1}{2}$ has the same law as $k_{\max} = \max_{k \in S} k$, the rightmost occupied site observed in the ground state of $H_{\theta \gamma}$.

The Schur measures comprise an infinite parameter family generalising the Poissonised Plancherel measure appearing in the longest increasing sequence problem. Along with their time-dependant extensions~\cite{Okounkov_Reshetikhin_2003}, they encode a number of combinatorial models such as dimer tilings~\cite{BBCCR_2015} and last passage percolation~\cite{Johansson_2000}. 

\paragraph*{(ii) Unitary matrix models:} If we consider the cumulative distribution of the rightmost occupied site $k_{\max}$ it  can be expressed as a unitary matrix integral: for each positive integer $\ell$ we have 
\begin{equation}
 e^{\theta^2 \sum_{r}  r \gamma_r^2} \P_{\theta \gamma}(k_{\max} < \ell) = \int_{\mathcal{U}(\ell)} e^{\theta \tr \sum_{r \geq 1} (-1)^{r-1} \gamma_r (U^r + U^{-r})} \mathcal{D } U
\end{equation}
where $ \mathcal{D } U$ denotes the Haar measure on the unitary group $\mathcal{U}(\ell)$. This integral may naturally be interpreted as the partition function $Z_\ell$ for a probability density $p_{\theta \gamma;\ell}(U) = Z_\ell^{-1}e^{\theta \tr \sum_{r \geq 1} (-1)^{r-1} \gamma_r (U^r + U^{-r})}$ on the $\ell \times \ell$ unitary matrices. The density recovered for $\gamma_1 = 1$, $\gamma_{r>1} = 0$ was studied by Gross, Witten~\cite{Gross_Witten_1980} and Wadia~\cite{Wadia_1980} in the context of lattice gauge theory, who notably showed that as $\ell \to \infty$ with $\theta \sim \ell/x$, this model exhibits a third order ``strong-to-weak coupling'' phase transition at $x =2$. The Gross--Witten--Wadia model correspondence has been used to study fermionic systems notably in~\cite{Pallister_2022,Perez-Garcia_Tierz_2014}. Solutions of matrix models with more general densities $p_{\theta \gamma;\ell}$ were studied in~\cite{Jurkiewicz_Zalewski_1983}, and Periwal and Shevitz~\cite{Periwal_Shevitz_1990,Periwal_Shevitz_1990_2} showed that the coefficients $\gamma_r$ could be tuned to have a ``multicritical'' third order phase transition with scaling exponent $2+\tfrac{1}{m}$ for any positive integer $m$, generalising the $m=1$ case.  

\paragraph*{(iii) Momentum space trapped fermion models:} Consider $H_{\theta\gamma}$ as $\theta$ tends to infinity, with the scalar term removed for convenience. Looking at a small window around a lattice site, we can identify  suitable coefficients $b, d$ and critical exponent $q \in (0,1)$ such that scaling the site labels as $k \sim b\theta + x(d \theta)^{q}$ recovers a meaningful limiting Hamiltonian in terms of a continuous coordinate $x$ (on the assumption that discrete difference operators can be approximated by derivatives).  If $m$ is the smallest positive integer such that $\sum_{r} r^{2m+1}\gamma_r \neq 0$, upon setting $b = 2 \sum_r r \gamma_r $, $d = \frac{2(-1)^{m+1}}{(2m)!}\sum_{r} r^{2m+1}\gamma_r$ and $q = \frac{1}{2m+1}$ we have the heuristic
\begin{equation} \label{eq:edgeheuristic}
{H}_{\theta\gamma} = (d\theta)^{\frac{1}{2m+1}} \int \normord{c^\dagger_x \bigg[x + (-1)^m\frac{d^{2m}}{dx^{2m}}\bigg] c_x }\dd x + O(\theta^{-\frac{2m+2}{2m+1}})
\end{equation}
as $\theta \to \infty$ with $k \sim b\theta + x(d \theta)^{q}$ (see~\cite[Section 2.2]{BBW_2023} for details). The dominant term of $(d\theta)^{-\frac{1}{2m+1}} {H}_{\theta\gamma}$ is also recovered from models of fermions in continuous unidimensional space subject to \emph{flat trap} potentials $V(x) = x^{2m}$. In~\cite{LDMS_2018}, the authors considered these models in terms of a momentum coordinate $p$. The momentum density vanishes at the Fermi edge $p_F$, and the fluctuations around this point are described by rescaling the coordinate to $\tilde{p} = (p-p_{F}){p_F}^{\frac{1}{2m+1}}$ and linearising the kinetic term $\frac{1}{2}p^2$, to obtain a Hamiltonian of the above form. Its eigenfunctions are given by the order-$m$ Airy function defined at~\eqref{eq:Airyfnm}, since we have
\begin{equation}
\bigg(x + (-1)^m\frac{d^{2m}}{dx^{2m}} \bigg) \Ai_{2m+1}(x+v) = -v  \Ai_{2m+1}(x+v);
\end{equation}
hence, the authors find that the distribution of the maximum momentum $p_{\max}$ satisfies
\begin{equation}
\P\bigg( \frac{p_{\max} - p_F}{p_F^{-\frac{1}{2m+1}}} < s\bigg) \to F_{2m+1}(s)
\end{equation}
as the particle number and hence Fermi momentum tend to infinity.

\paragraph*{}Here we study the ground state of $H_{\theta \gamma}$ as $\theta$ tends to infinity directly, by finely analysing the \emph{correlation kernel} (or propagator)
\begin{equation}
K_{\theta \gamma}(k,\ell) = \langle c^\dagger_k c_\ell \rangle_{\theta \gamma}
\end{equation}
 where $\langle \cdot \rangle_{\theta \gamma}$ denotes averaging with respect to the ground state of $H_{\theta \gamma}$. In terms of the random half-integer sequence $S$, the kernel satisfies $ \P_{\theta \gamma}(X \subset S) = \det_{k,\ell \in X }K_{\theta\gamma}(k,\ell) $
for any finite sequence of half-integers $X$ (hence determining the probability of any measurement which is physically possible). A limit of $K_{\theta\gamma}(k,k)$ gives the model's \emph{limit density}, that is 
$ \varrho(x) := \lim_{\theta \to \infty } \P_{\theta\gamma}(\lfloor x \theta\rfloor - \tfrac{1}{2} \in S)$. The integral of this density is a ``limit shape'' of the model: if $N(k)$ denotes the number of occupied sites to the right of $k$, then $N(x\theta) / \theta $ converges uniformly in probability to $\int_x^\infty \varrho(x') \dd x'$.  Under the corresponding measures on partitions, the rescaled profile of the Young diagram has a limiting curve determined by this integral (see Appendix~\ref{app:schur}). 
By the inclusion-exclusion principle, the gap probabilities (i.e., the probabilities of observing no occupied sites in given intervals) of the model are given by Fredholm determinants of the kernel, so the asymptotic interface fluctuations may be extracted from a suitable scaling limit of $K_{\theta\gamma}$.

 As $\theta\to \infty$, the ground state of $H_{\theta\gamma}$ may be characterised by the function
\begin{equation} \label{eq:Dfn}
D(\phi) := \sum_{r \geq 1} 2 r \gamma_r \cos  r \phi,
\end{equation}
which is even and $2\pi$-periodic. The case where $D(\phi)$ is decreasing on the interval $[0,\pi]$ was previously treated in~\cite{BBW_2023} 
(many of the results reviewed here were found independently in~\cite{Kimura_Zahabi_2021} and announced in the short paper~\cite{FPSAC_preprint_BBW}, but without the same hypotheses; we primarily cite the former work for this reason). With that constraint on the coefficients $\gamma_r$, we have a straightforward expression for the limit density, with
\begin{equation}
\varrho(x) = \frac{\chi}{\pi}
\end{equation}
where $\chi = \chi(x)$ is the (unique) solution of $D(\chi) =x$ in $[0,\pi]$ for each $x$ in the range of $D$, $\chi = \pi$ for $x> b = \max D$ and $\chi= 0$ for $x<-\tilde{b} = \min D$ (explicitly, $b = 2 \sum_r r \gamma_r$ and $\tilde{b} = 2 \sum_r (-1)^{r+1} r \gamma_r$). The range of $D$ corresponds to the \emph{bulk} of the model, to the left of the bulk we have \emph{frozen region} and to the right of it we have the \emph{empty region} with $\varrho =0$.  The local limit of $K_{\theta\gamma}$ is a discrete sine kernel, with 
\begin{equation}
K_{\theta\gamma}(\lfloor x\theta \rfloor + s - \tfrac{1}{2}, \lfloor x\theta \rfloor + t - \tfrac{1}{2}) \to 
\frac{\sin \chi (s-t)}{\pi (s-t)}
\end{equation}
as $\theta \to \infty$ for distinct fixed integers $s, t$ (note that it vanishes in the frozen and empty regions).

 At the right edge of the bulk, the limit density vanishes as
\begin{equation} \label{eq:lsedge}
\varrho(x) \sim \frac{1}{\pi}\bigg(\frac{b-x}{d}\bigg)^{\frac{1}{2m}} \qquad \text{as } x\to b^-.
\end{equation}
where $m$ is the smallest positive integer such that $\sum_{r} r^{2m+1}\gamma_r \neq 0$ and $d = 2(-1)^{m+1}\sum_{r} r^{2m+1}\gamma_r/(2m)!$. In a window scaling critically with $\theta^{\frac{1}{2m+1}}$ around $b\theta$, the kernel has a universal limit for each $m$, given by order $m$ Airy kernel defined at~\eqref{eq:Airyorderm}. We have
\begin{equation} \label{eq:onecutedgelimit}
(d\theta)^{\frac{1}{2m+1}}K_{\theta\gamma}(\lfloor b\theta + x(d\theta)^{\frac{1}{2m+1}}\rfloor - \tfrac{1}{2}, \lfloor b\theta + y(d\theta)^{\frac{1}{2m+1}}\rfloor - \tfrac{1}{2}) \to \mathcal{A}_{2m+1}(x,y)
\end{equation}
as $\theta \to \infty$ for all $x,y$ in compact subsets of $\R$. As a consequence, we find that in the same scaling regime, the rightmost occupied site asymptotically follows the order-$m$ Tracy--Widom distribution, with 
\begin{equation} \label{eq:singlefermidist}
\lim_{\theta \to \infty} \P_{\theta\gamma}\bigg(\frac{ k_{\max} - b\theta}{(d\theta)^{\frac{1}{2m+1}}} <s \bigg) = F_{2m+1}(s) := \det(1-\mathcal{A}_{2m+1})_{L^2([s,\infty))}. 
\end{equation} 
This is precisely what we would predict from the heuristic correspondence between $H_{\theta\gamma}$ and the momentum space flat trap Hamiltonian of~\cite{LDMS_2018} in the edge scaling limit. Note, however, that the hypothesis that $D(\phi)$ is decreasing on $[0,\pi]$ does not follow directly from the heuristic. 

The convergence to $F_{2m+1}$ can equally be written for the partition function of a model of $\ell\times\ell$ random unitary matrices with density $p_{\theta \gamma,\ell}$ in a regime where $\ell \sim b \theta + s(d\theta)^{\frac{1}{2m+1}}$. 
Combining this result with~\cite[Theorem~1.1]{Cafasso_Claeys_Girotti_2019}, this notably shows that the densities $p_{\theta \gamma,\ell}$ define  multicritical unitary matrix models in the sense of Periwal and Shevitz~\cite{Periwal_Shevitz_1990,Periwal_Shevitz_1990_2}, with the same phase transition  at $x=b$ a regime where  $\ell \sim x\theta$ as $\theta \to \infty$; see~\cite{Kimura_Zahabi_2-2021b} for a detailed discussion.
As shown in~\cite{BBW_2023}, as $x \to b^+$ a cut appears in the support of the limiting eigenvalue density $\rho$, with the density vanishing as 
\begin{equation}
\rho(e^{i\alpha}) \sim \frac{1}{2\pi} \frac{d}{b}(\pi - \alpha)^{2m} \qquad \text{as }\alpha \to \pi^-. 
\end{equation}

\subsection{Main results}\label{sec:main_results}
We extend our study of the ground state of $H_{\theta \gamma}$ to the case where $D$ is not monotonous on $[0,\pi]$. Here, the limit density and the edge behaviour depends on the nature of the \emph{Fermi seas} associated with a given choice of coefficients $\gamma$.

\paragraph*{The Fermi sea} For each $x$, let $0\leq  {\chi}_1 < {\chi}_2 < \ldots < {\chi}_{2p} \leq \pi$ be the angles such that for all $\phi$
 in the set 
 \begin{equation}
 I_x := [-{\chi}_{2p} , -{\chi}_{2p-1}] \cup  \ldots \cup[-{\chi}_2 , -{\chi}_1]\cup [{\chi}_1 , {\chi}_2] \cup  \ldots \cup [{\chi}_{2p-1} , {\chi}_{2p}] 
 \end{equation} 
  we have $D(\phi) - x\geq 0$, and for all $\phi \in [\pi, -\pi] \setminus I_x$ we have $D(\phi) - x <0$. The domain $I_x$ is called the {Fermi sea} of the model at $x$, as it represents the quantum numbers associated with a site $k \sim x\theta$ as $\theta \to \infty$ in the ground state. We say $I_x$ has $n$ cuts if there are $n$ gaps in the subset $\{e^{i\phi}, \phi \in I_x\}$ on the unit circle; note that if $\chi_1 = 0$ or is $\chi_{2p} = \pi$, they do not correspond to cuts in $I_x$, and for each cut we have two distinct solutions to $D(\chi_i) - x = 0$. For $x > b = \max D$, the Fermi sea $I_x$ is just the empty set, and for $x < -\tilde{b} = \min D$ we have $I_{x} = [-\pi,\pi]$. 

\begin{figure}

{\def\svgwidth{0.97\textwidth} 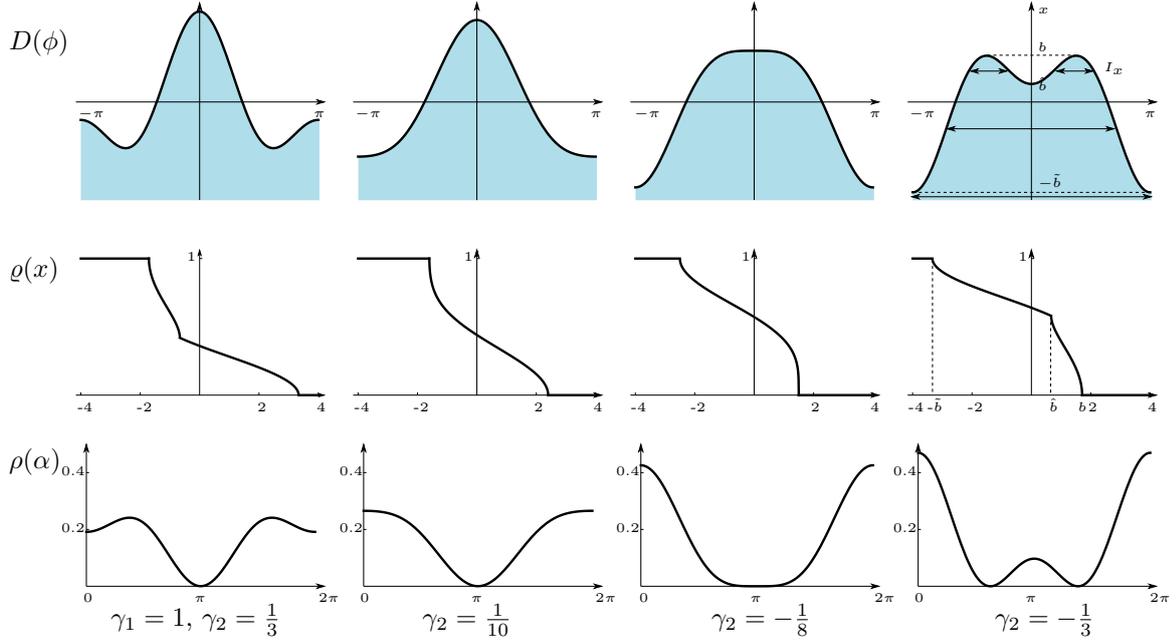 }
\caption{Plots of the function $D$ determining the Fermi seas (top), the limit fermion density $\varrho$ (middle) and corresponding limiting eigenvalue density $\rho$ in the $\theta \sim \ell/b$ regime for four sets of coefficients $\gamma$ with $\gamma_1 = 1$ and $\gamma_r = 0$ for $r >2$. For $\gamma_2 = \tfrac{1}{10}$ and $\gamma_2 = -\tfrac{1}{8}$ (illustrated in the central columns), the Fermi sea has one-cut throughout and the limit density is smooth in the bulk; the $\gamma_2 = -\tfrac{1}{8}$ model is order $m=2$ multicritical. For $\gamma_2 = \pm \tfrac{1}{3}$ (the leftmost and rightmost columns), the Fermi sea splits into two in the bulk, and there is a discontinuity in the derivative of the limit density where it splits. In the $\gamma_2=\frac{1}{3}$ case, there are two cuts in the Fermi sea $I_{b^-}$ before the right edge. Note that in this case the limiting eigenvalue density goes to $0$ at two points. } \label{fig:4casesDetc}
\end{figure}

 \paragraph*{Limit density and local limiting kernel} At each point $x$, the limit density may be written in terms of the boundaries $\chi_i$ of the $I_x$ as
\begin{equation} \label{eq:ls}
\varrho(x) =  \frac{{\chi}_{2p}}{\pi} -  \frac{{\chi}_{2p-1}}{\pi} + \ldots + \frac{{\chi}_{2}}{\pi}- \frac{{\chi}_{1}}{\pi}.
\end{equation}
See Figure~\ref{fig:4casesDetc} for some limit densities and corresponding Fermi seas. 
In the local limit, the ground state propagator converges to an extended sine kernel, with 
\begin{equation}
K_{\theta\gamma}(\lfloor x\theta \rfloor + s - \tfrac{1}{2}, \lfloor x\theta \rfloor + t - \tfrac{1}{2}) \to \frac{\sin {\chi}_{2p} (s-t)}{\pi (s-t)} - \frac{\sin {\chi}_{2p-1} (s-t)}{\pi (s-t)} + \ldots -\frac{\sin {\chi}_{1}(s-t)}{\pi (s-t)}
\end{equation}
as $\theta \to \infty $ for fixed integers $s \neq t$. We prove this in Section~\ref{sec:lsfromk}. This is a generalisation from the case where $D$ is decreasing on $[0,\pi]$, for which  $\chi_1 = 0$ everywhere and the only cut in $I_x$ is bounded by $\pm\chi_2$. Again, we have a frozen region with $\varrho(x) = 1$ for $x < -\tilde{b}$ and an empty region with $\varrho(x) = 0$ for $x>b$.
In the case where $I_x$ has multiple cuts, the local ground state near $x\theta$ is analogous to the ``Moses states'' studied by Fokkema, Eliëns and Caux~\cite{Fokkema_Eliens_Caux_2014} in the context of excited states of integrable models. 
If the Fermi sea splits within the bulk, we find a rather more exotic limit density than in the case where $D$ is monotone on $[0,\pi]$, as we have a discontinuity in the derivative of $\varrho$ at any point where the number of cuts in $I_x$ changes (see Figure~\ref{fig:4casesDetc}). In this sense, these models can have multiple interfaces, but we leave them for future work and concentrate on the interface for which we might expect universal behaviour.

\paragraph*{Edge scaling limit and fluctuations in the two-cut case} The main result of this work concerns the behaviour of fermions near the edge of the limit shape, at $b:= \max D$, which we find depends on the number of cuts in $I_{b^-}$, the Fermi sea at $x$ immediately below $b$. In the case where $I_{b^-}$ has a single cut, we have the same ``universal'' convergence to order-$m$ Airy kernel of the form~\eqref{eq:onecutedgelimit} in an appropriate scaling limit, along with universal asymptotic fluctuations governed by $F_{2m+1}$. However, if $I_{b^-}$ has multiple cuts, the universality breaks down and we do not in general  find a true limiting kernel. We consider the case where $I_{b^-}$ has two cuts and does not contain $0$ (see an example in the top right corner of Figure~\ref{fig:4casesDetc}), and let $m$ and $d$ be the constants such that the limit density vanishes as
\begin{equation}
\varrho(x) \sim \frac{2}{\pi} \bigg(\frac{b-x}{d}\bigg)^{\frac{1}{2m}} \qquad \text{as }x \to b^-.
\end{equation}
We show that, in a regime where $k \sim b\theta + x(d\theta)^{\frac{1}{2m+1}}$ and $\ell \sim b\theta + y(d\theta)^{\frac{1}{2m+1}}$,  $K_{\theta \gamma}(k,\ell)$ tends to the order-$m$ Airy kernel $\mathcal{A}_{2m+1}(x,y)$ multiplied by a rapidly oscillating bounded factor of $2\cos \chi_b (d\theta)^{\frac{1}{2m+1}}(x-y)$ (see Section~\ref{sec:edgekernel}). We argue that oscillation terms of this kind appear also appear when $I_{b^-}$ has more than two cuts, and are a consequence of the discrete nature of the fermion model. A similar phenomenon was observed by Bonnet, David and Eynard~\cite{Bonnet_David_Eynard_2000} in multi-cut Hermitian matrix models, where the two-point correlation function exhibited oscillations due to the discreteness of the number of eigenvalues (these models were more recently studied rigorously by Borot and Guionnet~\cite{Borot_Guionnet}). 

We argue that in the case where $I_{b^-}$ has two cuts, the asymptotic  edge fluctuations are governed by a new asymptotic distribution. At the same scale of $\theta^{\frac{1}{2m+1}}$ predicted by the vanishing exponent from the limit density, by averaging over the oscillating factor we obtain 
\begin{equation}\label{eq:twocutfluctuations}
\lim_{\theta \to \infty}\P_{\theta\gamma}\bigg( \frac{k_{\max} - b\theta}{(d\theta)^{\frac{1}{2m+1}}} < s\bigg) = \det(1-\mathcal{A}_{2m+1})_{L^2([s,\infty))}^2 = F_{2m+1}^2(s), 
\end{equation} 
that is, the square of the order $m$ Tracy--Widom distribution (see Section~\ref{sec:edgefluctuation}). An analogous limiting distribution $F_{2m+1}^n$ can be recovered when $I_{b^-}$ has $n$ cuts. These distributions have a natural interpretation as the distribution of the maximum element of $n$independant copies of the order-$m$ Airy process obtained in the limit in the case where $I_{b^-}$ has one cut. We conjecture that the limiting point processes in these cases can be described by $\alpha$-determinantal point process, which have notably been found by Cunden, Majumdar and O'Connell~\cite{Cunden_2019} for fermions in a harmonic potential in certain excited states. 

\paragraph*{Correspondence with multi-cut unitary matrix models} The estimate~\eqref{eq:twocutfluctuations} for the asymptotic fluctuations in $k_{\max}$ in the two-cut also applies to the partition function of the corresponding unitary matrix model with density $p_{\theta\gamma, \ell}$ in the $\ell \sim b\theta + s(d\theta)^{\frac{1}{2m+1}}$ regime. Where $\ell \sim x \theta$, this model has a phase transition at $x=b$.  We show, informally, that in the case where the Fermi sea $I_{b^-}$ has $n$ cuts, $n$ cuts appear in the support of the limiting eigenvalue density as $x \to b^+$ (see the bottom right of Figure~\ref{fig:4casesDetc} for an example, and Section~\ref{sec:matrixmodels} for calculations).

\subsection{Outline} In section~\ref{sec:bulk}, we consider the ``macroscopic'' properties of the ground state of $H_{\theta\gamma}$ as $\theta \to \infty$, by finding the limit density and the local limiting kernel. We first present a heuristic approach, using a local density approximation to identify the Fermi seas of the model, then study the kernel by rigorous saddle point analysis. In a simple case with a two-cut phase, we give explicit Fermi seas. In Section~\ref{sec:edge}, we look at the behaviour in a scaling liming near the vanishing edge in the case where the Fermi sea has two cuts. We find a rapidly oscillating ``limiting'' kernel, and estimate the asymptotic distribution for the fluctuations. Finally in Section~\ref{sec:matrixmodels} we review the correspondence with unitary matrix models. We find the limiting distribution of eigenvalues on the unit circle for a density $p_{\theta\gamma,\ell}$ in a weakly interacting regime where $\ell / \theta > b$, and relate the cuts appearing in its support as $x \to b^+$ to the cuts in the Fermi sea. We refer only to the lattice fermion picture in the main text, but in Appendix~\ref{app:schur} we recall the correspondence with Hermitian Schur measures and restate our results in those terms. 

\section{Limit densities and local limiting kernels} \label{sec:bulk}

\subsection{Local density approximation and Fermi seas}  \label{sec:lda}
Let us consider the kernel $K_{\theta\gamma}(k,\ell)$ as $\theta \to \infty$ where $k\sim x\theta + \delta$ and $\ell \sim x\theta + \epsilon$ for $\delta,\epsilon$ which we assume to be very small relative to system size, but large relative to the distance between sites. Then, we may make a \emph{local density approximation} (LDA; see e.g.~\cite{Stephan_2019}), and assume that as $\theta\to\infty$ the sites $k,\ell$ only ``see'' a homogeneous system, with a constant potential. Switching to Fourier space, by transforming  the fermionic creation operator to $\hat{c}^\dagger(\phi) := \sum_{k} e^{i k \phi} c^\dagger_k$, diagonalises the Hamiltonian; assuming the potential behaves as $k \sim x \theta$  as $\theta \to \infty$, we have
\begin{equation}
H_{\theta\gamma}  \approx   \frac{\theta}{2\pi}\int_{-\pi}^\pi \big(x - \sum_{r\geq 1}2 r \gamma_r \cos r \phi\big) c^\dagger(\phi)c(\phi)\dd\phi
\end{equation}
The ground state corresponds to the Fourier frequencies with dispersion $(x - \sum_{r}2 r \gamma_r \cos r \phi) \leq 0$. This defines the Fermi sea, and it corresponds to the set $ I_x$ defined in Section~\ref{sec:main_results} above. Under the LDA, the kernel is simply a projector onto $ I_x$; as $\theta \to \infty$ we have 
\begin{align}
K_{\theta\gamma}(k,\ell) = \langle c_k^\dagger c_\ell \rangle_{\theta\gamma} &\approx \frac{1}{2\pi}\int_{ I_x }  e^{i\phi (\delta - \epsilon)} \dd \phi; \\
&= \frac{1}{2\pi}\int_{-{\chi}_{2p}}^{{\chi}_{2p}}  e^{i\phi (\delta - \epsilon)} \dd \phi - \frac{1}{2\pi}\int_{-{\chi}_{2p-1}}^{{\chi}_{2p-1}}  e^{i\phi (\delta - \epsilon)} \dd \phi  + \ldots - \frac{1}{2\pi}  \int_{-{\chi}_1}^{{\chi}_1}e^{i\phi (\delta - \epsilon)} \dd \phi 
\end{align}
where the angles $\{{\chi}_i\}$ determining the boundaries of the disjoint intervals in $I_x$. This gives
\begin{equation}
K_{\theta\gamma}(k,\ell) \approx \frac{\sin {\chi}_1 (\delta - \epsilon)}{\pi (\delta - \epsilon)} - \frac{\sin {\chi}_2 (\delta - \epsilon)}{\pi (\delta - \epsilon)} + \ldots + \frac{\sin {\chi}_{2p-1}(\delta - \epsilon)}{\pi (\delta - \epsilon)}-\frac{\sin {\chi}_{2p}(\delta - \epsilon)}{\pi (\delta - \epsilon)},
\end{equation}
as $\theta \to \infty$, 
giving the local limit of the kernel under the assumption that this is the appropriate scale for the LDA. Setting $\delta = \epsilon$, we obtain the limit density~\eqref{eq:ls}.

\subsection{Exact kernel and its local limit}  \label{sec:lsfromk}
In this section we re-derive the limit density and local limiting kernel without making any assumptions as to the scales associated with the model, by rigorous saddle point analysis of an exact contour integral expression for $K_{\theta\gamma}$. This expression was found in~\cite{Okounkov_2001}, we reproduce a derivation in more physical notation. The procedure we use to analyse it is directly adapted from works of Okounkov and coauthors, detailed for instance in~\cite{Okounkov_2003} (see e.g.~\cite[Chapter~VIII]{Flajolet_Sedgewick_2009} for a general introduction to saddle point methods). This approach recovers the same local limiting kernel as the LDA, and additionally allows us to see the scale of fluctuations. 

\paragraph*{Integral expression for the kernel}  Consider the Hamiltonian $H_0 := \sum_{k \in \Z+\frac12} k \normord{c^\dagger_kc_k}$, with no hopping terms. Its ground state is the \emph{domain wall} state $\ket{\emptyset}$ with all negative sites occupied and all positive sites vacant. Then, we introduce the operator
\begin{equation}
a_r := \sum_{k \in \Z+\frac{1}{2}} \normord{c^\dagger_k c_{k+r}}
\end{equation}
for each positive integer $r$, along with its adjoint $a_r^\dagger = a_{-r}$; from the fermionic canonical anticommutation relations
\begin{equation}
 \{c_k,c^\dagger_\ell\} := c_kc^\dagger_\ell -c^\dagger_\ell c_k =\delta_{k,\ell}, \qquad \{c_k,c_\ell\} = \{c^\dagger_k,c^\dagger_\ell\} = 0,
 \end{equation} 
we have that the $a_r$ satisfy the bosonic canonical commutation relation
\begin{equation}
[a_r,a_s^\dagger] = a_r a_s^\dagger - a_s^\dagger a_r =  r \delta_{r,s}
\end{equation}
along with $[H_0,a_r] = -ra_r$ and $[H_0,a^\dagger_r] = ra^\dagger_r$. From the formula 
\begin{equation} \label{eq:bch}
e^{A}Be^{-A} = \sum_{n=0}^\infty \frac{1}{n!} \underbrace{[A,[A,\ldots[A}_{n\text{ times}},B]\ldots]] ,
\end{equation} for a given sequence of coefficients $\gamma$ we have
\begin{equation}
H_{\theta\gamma} = H_0 - \theta \sum_{r \geq 1} r \gamma_r (a_r+a_{r}^\dagger) + \theta^2 \sum_{r\geq 1}  r^2 \gamma_r^2 =  \mathcal{U}_{\theta\gamma} H_0 \mathcal{U}_{\theta\gamma}^{-1} 
\end{equation} 
in terms of the unitary operator
\begin{equation} \label{eq:unitaryop}
\mathcal{U}_{\theta\gamma} := e^{ \theta \sum_r  \gamma_r (a_r^\dagger - a_r)}. 
\end{equation}
Then we can write the ground state of $H_{\theta\gamma}$ in terms of the domain wall state as $\mathcal{U}_{\theta\gamma} \ket{\emptyset}$.

From this, we can find a generating function for the kernel. Setting $c^\dagger(z) := \sum_{k} z^k c^\dagger_k$ and $c(w) := \sum_{k} w^{-k} c_k$, we have 
\begin{equation} \label{eq:accomms}
[a_r,c^\dagger(z)] = z^rc^\dagger(z)\quad \text{and} \quad [a_r,c(w)] = -w^r c(w),
\end{equation}
and by applying~\eqref{eq:bch} we have
\begin{equation} \label{eq:Ucommutators}
\mathcal{U}_{\theta\gamma} c^\dagger(z) = e^{\theta \sum_r \gamma_r (z^{r}-z^{-r})} c^\dagger(z) \mathcal{U}_{\theta\gamma} \quad \text{and} \quad \mathcal{U}_{\theta\gamma} c(w) = e^{\theta \sum_r \gamma_r (w^{r}-w^{-r})} c(w) \mathcal{U}_{\theta\gamma}.
\end{equation}
Then, the generating function of the kernel is
\begin{align}
\sum_{k,\ell \in \Z+\frac12} z^kw^{-\ell} K_{\theta\gamma}(k,\ell) 
& = \bra{\emptyset} \mathcal{U}^{-1}_{\theta\gamma} c^\dagger(z)c(w) \mathcal{U}_{\theta\gamma} \ket{\emptyset} \\
&= e^{\theta \sum_r \gamma_r (z^r-z^{-r})} \bra{\emptyset} c^\dagger(z)c(w) \ket{\emptyset} e^{\theta \sum_r \gamma_r (w^{-r}-w^{r})};
\end{align}
evaluating the final expectation on the domain wall state, we have
\begin{equation}
\bra{\emptyset} c^\dagger(z)c(w) \ket{\emptyset} = \sum_{k,\ell \in \Z+\frac12} \frac{z^k}{w^{\ell}} \delta_{k\ell} \ind_{k<0} = \frac{\sqrt{zw}}{z-w} \qquad \text{for } |w|<|z|.
\end{equation}
The kernel is extracted from this generating function with the double contour integral
\begin{equation} \label{eq:kexact}
K(k,\ell) = \frac{1}{(2\pi i)^2}\oiint_{c_+,c_-} \frac{e^{\theta \sum_r \gamma_r (z^r-z^{-r})}}{e^{\theta \sum_r \gamma_r (w^{r}-w^{-r})}} \frac{\sqrt{zw}}{z-w} \frac{\dd z \dd w}{z^{k+1} w^{-\ell+1}}
\end{equation}
where the integral in $w$ is taken over a contour $c_-$ running counter-clockwise around the origin and the integral in $z$ is over a contour $c_+$ enclosing $c_-$.  

\begin{figure}

{\def\svgwidth{\textwidth} \input{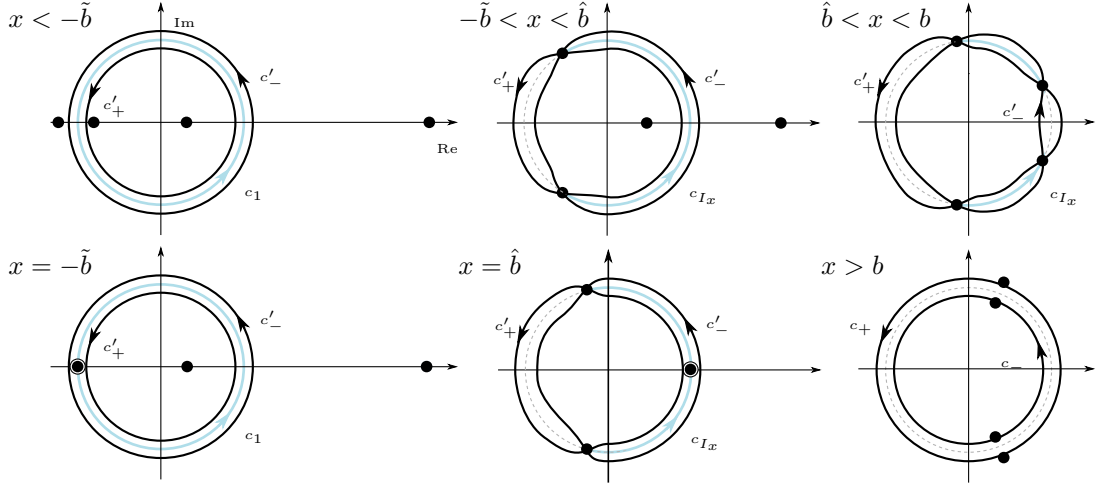} }
\caption{Saddle points of the action $S(z;x)$ for $\gamma_1 = 1$, $\gamma_2 = -\tfrac13$ and $\gamma_r = 0$ for $r>2$ (simple saddle points are correspond to single black dots, double saddle points correspond to encircled black dots), along with  the appropriate integration contours. The frozen region corresponds to $x \leq-\tilde{b} = -\tfrac{10}{3}$ where we have $I_x = [-\pi,\pi]$. The appropriate integration contours are shown on the left, with the saddle points for $x= \tfrac72$ above and $x=-\tfrac{10}{3}$ below. For $-\tilde{b}<x\leq \hat{b}=\tfrac{2}{3}$, $I_x$ has a single cut and contains $0$, and the contours have the form shown in the central column (with saddle points for $x=-\tfrac23$ above and $x=\frac23$ below, where there is a double saddle point at $1$). For $\hat{b}<x< b = \tfrac{41}{24}$, $I_x$ has two cuts, and we use contours of the kind show in the top right. The empty region corresponds to $x\geq b$, and we use the undeformed contours $c_+,c_-$ shown in the bottom right (with the saddle points for $x = \tfrac74$). The corresponding $D$ and limit density $\varrho$ are shown on the right of Figure~\ref{fig:4casesDetc}.} \label{fig:intcontours}
\end{figure}

\paragraph*{Saddle point analysis of the kernel} Let us introduce some useful notation. The \emph{action} associated with the kernel $K_{\theta\gamma}$ is
\begin{equation} \label{eq:action}
S(z;x) = \sum_{r\leq 1} \gamma_r(z^r - z^{-r}) - x \log z,
\end{equation}
so that we have
\begin{equation} \label{eq:kdoubleint}
K(\lfloor x \theta\rfloor+s -\tfrac12,\lfloor x \theta\rfloor+t -\tfrac12) = \frac{1}{(2\pi i)^2}\oiint_{c_+,c_-} \frac{e^{\theta [S(z;x) - S(w;x)]}}{z^{s+\frac12}w^{-t+\frac12}(z-w)}  [1+o(1)]\dd z \dd w
\end{equation}
as $\theta \to \infty$ (the $o(1)$ accounts for the difference between $x\theta$ and its integer part, and is uniform).
To begin with, we take the contours to be
\begin{equation}
 c_{\pm} = \{ (1 \pm \epsilon) e^{i\phi}, \phi \in [-\pi,\pi]\}
 \end{equation} for small $\epsilon>0$, passing either side of the unit circle $c_1$. If $S(z;x)$ has saddle points along $c_1$, they are at $z = e^{i \phi}$ where $\phi$ are the (real) solutions of 
\begin{equation} \label{eq:c1saddlepoints}
z\frac{\dd}{\dd z} S(z;x)\vert_{z = e^{i\phi}} = D(\phi) - x = 0
\end{equation}
where $D(\phi)$ is the function  defined at~\eqref{eq:Dfn}. Let $b = \max D$ and let $-\tilde{b} = \min D$ (both $b$ and $\tilde{b}$ are positive). We consider the asymptotics in three phases. See Figure~\ref{fig:intcontours} for examples of the contours used in each phase.

\noindent{\sffamily\textbf{(i)}} For $x> b$,~\eqref{eq:c1saddlepoints} has no solutions and there are no saddle points on the unit circle. 
Along the undeformed contour $c_+$, we have
\begin{equation} \label{eq:reSbound1}
\Re\big[S(z;x)\vert_{z = (1+\epsilon)e^{i\phi}}\big] = \epsilon (D(\phi) -x) + O(\epsilon^2),
\end{equation}
and the dominant term is negative for all $\phi$. Similarly, on $c_-$ we have
\begin{equation} \label{eq:reSbound2}
\Re\big[S(w;x)\vert_{w = (1-\epsilon)e^{i\phi}}\big] = -\epsilon (D(\phi) -x) + O(\epsilon^2),
\end{equation}
with a positive dominant term. Hence, for $\epsilon$ sufficiently small, we have  $e^{\Re[S(z;x)-S(w;x)]} < e^{-C\theta}$ for a positive constant $C$, for all $(z,w) \in c_+ \times c_-$. Since $|z-w| \geq 2 \epsilon$ along these contours, every other term is bounded and we have overall exponential decay. This in turn implies dominated convergence, so the integral also decays and 
\begin{equation} \label{eq:reSbound1}
\lim_{\theta \to \infty } K_{\theta \gamma}(\lfloor x \theta\rfloor+s -\tfrac12,\lfloor x \theta\rfloor+t -\tfrac12) = 0
\end{equation}
(this is the empty region).

\noindent {\sffamily\textbf{(ii)}} Now consider a point $x \in (-\tilde{b},b)$, for which~\eqref{eq:c1saddlepoints} has at least two real solutions. Suppose $S(z;x)$ only has \emph{simple} saddle points on the unit circle, i.e. for each $\chi$ satisfying $D(\chi) - x = 0$, we have $D^\prime(\chi) \neq 0$. Then, each of the saddle points on the unit circle is at a point $z = e^{\pm i\chi_i}$ where $\chi_i$ is a  boundary of one the intervals composing the Fermi sea $I_x$. Let $\partial I_x = \{\pm \chi_i\} \setminus \{0,-\pi,\pi\}$ denote the set of boundaries of $I_x$ within $(-\pi,\pi)$, and consider the deformed contours
\begin{equation}
c_\pm^\prime =\big\{ \big(1\pm \epsilon(\phi)\big)e^{i\phi}, \phi \in [-\pi,\pi] \setminus I_x \big\} 
\end{equation}
where $\epsilon$ is a real, continuous, $2\pi$-periodic and close to $0$ everywhere, and satisfies $\epsilon(\phi) >0$ for all $\phi \in [-\pi,\pi] \setminus I_x$, $\epsilon(\phi) = 0$ for all $\phi \in \partial I_x$ and $\epsilon(\phi) < 0 $ for all $\phi \in  I_x \setminus \partial I_x$.
These contours pass through the saddle points, and recalling~\eqref{eq:reSbound1} and~\eqref{eq:reSbound2} we have 
\begin{equation}
\Re[S(z;x) - S(w;x)] \leq 0 \qquad \text{ for all } (z,w) \in c_+^\prime \times c_-^\prime
\end{equation}
(which is an equality only for $z,w$ at saddle points). In order to write the kernel~\eqref{eq:kdoubleint} in terms of an integral over these contours, we need to deform the original contours $c_\pm$ over one another, across the subset 
\begin{equation}
c_{I_x} = \{e^{i\phi}, \phi \in I_x\}
\end{equation}
of the unit circle. In doing so we pick up the pole at $z-w$, which has a residue of $1/w^{s-t+1}$, and we have
\begin{equation} \label{eq:kbulkint}
K(\lfloor x \theta\rfloor+s -\tfrac12,\lfloor x \theta\rfloor+t -\tfrac12) = \frac{1}{2\pi i }\int_{c_{I_x}} \frac{\dd w}{w^{s-t+1}}  +\frac{1}{(2\pi i)^2}\oiint_{c_+^\prime,c_-^\prime} \frac{e^{\theta [S(z;x) - S(w;x)]}}{z^{s+\frac12}w^{-t+\frac12}(z-w)}  [1+o(1)]\dd z \dd w.
\end{equation}

First, let us consider the integral over $c^\prime_\pm$ as $\theta \to \infty$. These contours encounter the  $z = w$ singularity at each saddle point. Let us consider the integral over portions of $c_+'$ and $c_-'$ of length $2\delta$ around a saddle point $e^{i \chi}$, and the corresponding part for the conjugate saddle point $e^{-i\chi}$. Let us parametrise the argument of $z$ by $\phi = \pm\chi + \theta^{-\frac12}\zeta$, and the argument of $w$ by $\phi = \pm \chi + \theta^{-\frac12}\omega$.
 Then, we see that the sum of each of these parts has a dominant term of the form
 \begin{equation*}
 \frac{\theta^{-\frac{1}{2}}\cos\chi(s-t)}{(2\pi)^2} \iint_{[-\theta^{\frac{1}{2}} \delta, \theta^{\frac{1}{2}} \delta]^2}  \frac{e^{-C(\zeta^2 + \omega^2)} }{\zeta-\omega} \dd \zeta \dd w
 \end{equation*}
 for large $\theta$ and small $\epsilon$, where $C$ is a positive constant. 
 The final integral is finite;  noting that the finite number of other saddle points only contribute terms of this order in $\theta$, we see that the integral over $c_+' \times c_-'$ is $O(\theta^{-\frac{1}{2}})$.  Let us mention how this generalises for points $x \in (-\tilde{b},b)$ at which there are \emph{higher-order} saddle points (for instance, at points where the Fermi sea splits). If $\chi$ is a solution to $D(\chi) - x = 0$ and  $n$ is the smallest number for which we have $D^{(n)}(\chi) \neq 0$, we can repeat the same arguments the argument of $z$ parametrised by $\phi = \pm \chi + \theta^{-\frac{1}{n+1}}\zeta$ and $w$ parametrised similarly, and conclude that the contribution is $O(\theta^{-\frac{1}{n+1}})$. If there are no lower order saddle points on the unit circle, the integral over $c_+'\times c_-'$ is  $O (\theta^{-\frac{1}{n+1}})$. 

 Thus, the local limiting kernel is simply given by the first integral in~\eqref{eq:kbulkint}, which gives
\begin{align}
\lim_{\theta \to \infty} K(\lfloor x \theta\rfloor+s -\tfrac12,\lfloor x \theta\rfloor+t -\tfrac12) &= \frac{1}{2\pi i }\int_{c_{I_x}} \frac{\dd w }{w^{s-t+1}}  \\
&= \begin{cases} \frac{\chi_{2p}}{\pi} - \frac{\chi_{2p-1}}{\pi} + \ldots - \frac{\chi_{2}}{\pi} \qquad &\text{for }s=t,\\
\frac{\sin \chi_{2p}(s-t)}{\pi(s-t)} - \frac{\sin \chi_{2p-1}(s-t)}{\pi(s-t)} + \ldots - \frac{\sin \chi_{1}(s-t)}{\pi(s-t)} & \text{otherwise}. \notag
\end{cases}
\end{align}
Note that the rate of convergence is slower than in the empty region. The bulk has fluctuations at a scale of $\theta^{-\frac{1}{2}}$ at typical points, but we can tune the coefficients $\gamma_r$ to  have specific points with fluctuations at a smaller scale. 

\noindent {\sffamily\textbf{(iii)}} For $x<-\tilde{b}$, again there are no saddle points on the unit circle (recall that $D(\phi) -x>0$). Now we introduce the contours
\begin{equation}
c_\pm^\prime = \{(1\mp \epsilon)e^{i\phi}, \phi \in [-\pi,\pi]\}
\end{equation}
for small positive $\epsilon$; then by~\eqref{eq:reSbound1} and~\eqref{eq:reSbound2} we have that 
\begin{equation}
\Re[S(z;x) - S(w;x)] < 0 \qquad \text{for all } (z,w) \in c_+'\times c_-'.
\end{equation}
To deform $c_\pm$ to these contours, we pass them over  one another all along the unit circle $c_1$; from the $z=w$ residue we have
\begin{equation}
K(\lfloor x \theta\rfloor+s -\tfrac12,\lfloor x \theta\rfloor+t -\tfrac12) = \frac{1}{2\pi i }\oint_{c_{1}} \frac{\dd w}{w^{s-t+1}}  +\frac{1}{(2\pi i)^2}\oiint_{c_+^\prime,c_-^\prime} \frac{e^{\theta [S(z;x) - S(w;x)]}}{z^{s+\frac12}w^{-t+\frac12}(z-w)}  [1+o(1)]\dd z \dd w
\end{equation}
as $\theta\to \infty$. The integral over $c_\pm'$ decays exponentially as $\theta \to \infty$, and we have
\begin{equation}
 \lim_{\theta \to \infty}K(\lfloor x \theta\rfloor+s -\tfrac12,\lfloor x \theta\rfloor+t -\tfrac12) = \frac{1}{2\pi i }\oint_{c_{1}} \frac{\dd w}{w^{s-t+1}}  = \delta_{s,t},
 \end{equation} 
defining the frozen regime. Putting everything together we have the local limiting kernel in each regime, and the limit density $\varrho(x)$ from the $s=t$ case. 

\paragraph*{Convergence to a limit shape} The existence of the limit density $\varrho$ implies the emergence of a limit shape for the lattice fermion model: if we let $N(k)$ denote the random number of sites to the right of $k$ observed to be occupied in a measurement of the ground state of $H_{\theta\gamma}$, then 
\begin{equation} \label{eq:unifconv}
 \sup_{x \in \R} \bigg| \frac{N(x\theta)}{\theta }  - \int_x^\infty \varrho(x') \dd x' \bigg|  \xrightarrow{p} 0. 
 \end{equation} This follows from~\cite[Lemma~11]{BBW_2023}, which can be proved as follows. The expectation of $N(k)$ is simply the trace of $K_{\theta\gamma}$ from $k$ to infinity, and its variance is the trace of $ (K_{\theta\gamma} - K_{\theta\gamma}^2)$ from $k$ to infinity. This implies that, since the kernel is Hermitian, the variance of $N(k)$ is no greater than its expectation. The expectation of $N(x\theta)/\theta$ converges to the bounded function $\int_x^\infty \varrho(x') \dd x'$ as $\theta \to \infty$, so we have
 \begin{equation}
 \E_{\theta\gamma }  \bigg[\frac{N(x\theta)}{\theta}\bigg] \to \int_x^\infty \varrho(x') \dd x' \quad \text{and} \quad
 \Var \bigg[\frac{N(x\theta)}{\theta} \bigg] \leq \frac{1}{\theta} \E_{\theta\gamma } \bigg[ \frac{N(x\theta)}{\theta}\bigg] \to 0
 \end{equation}
as $\theta \to \infty$. Hence, $N(x\theta)/\theta$ converges pointwise in probability. Let $I$ be a bounded interval, and consider the finite set
 $I_\varepsilon = I \cap \varepsilon \Z$. From their definitions, both $N(x\theta)/\theta$ and $\int_x^\infty \varrho(x') \dd x'$ are 1-Lipschitz, so for any point $x$ in $ I$, there is a point $x_\varepsilon$ in $I_\varepsilon$ such that 
 \begin{equation}
\bigg| \frac{N(x\theta)}{\theta }  - \int_x^\infty \varrho(x') \dd x' \bigg| - \bigg| \frac{N(x_\varepsilon\theta)}{\theta }  - \int_{x_{\varepsilon}}^\infty \varrho(x') \dd x' \bigg| < \frac{\varepsilon}{2}.
 \end{equation}
 For any positive $\varepsilon$, we then have
 \begin{equation}
  \P_{\theta\gamma} \bigg(\sup_{x \in I} \bigg| \frac{N(x\theta)}{\theta }  - \int_x^\infty \varrho(x') \dd x' \bigg| > \varepsilon \bigg) \leq 
  \P_{\theta\gamma} \bigg(\sup_{x_{\varepsilon} \in I_{\varepsilon}} \bigg| \frac{N(x_{\varepsilon}\theta)}{\theta }  - \int_{x_{\varepsilon}}^\infty \varrho(x') \dd x' \bigg| > \frac{\varepsilon}{2} \bigg) \to 0
 \end{equation}
 as $\theta \to \infty$, by applying the pointwise convergence to the finite set $I_\varepsilon$. So, we have uniform convergence for $x$ on any bounded interval $I$, in particular we have it for $I = [-\tilde{b},b]$. For $x > b$, we have $\int_x^\infty \varrho(x') \dd x'  = 0$ and, by monotonicity,
 \begin{equation}
\sup_{x \in [b,\infty)}\bigg| \frac{N(x\theta)}{\theta }  - \int_x^\infty \varrho(x') \dd x' \bigg| = \sup_{x \in [b,\infty)} \frac{N(x\theta)}{\theta } = \frac{N(b\theta)}{\theta } \xrightarrow{p} 0
 \end{equation}
 as $\theta \to \infty$. Similarly, for $x<-\tilde{b}$ we have $\int_x^\infty \varrho(x') \dd x'  = -x$ and
  \begin{equation}
\sup_{x \in (\infty,-\tilde{b}]}\bigg| \frac{N(x\theta)}{\theta }  - \int_x^\infty \varrho(x') \dd x' \bigg| = \sup_{x \in (\infty,-\tilde{b}]} \bigg|  \frac{N(x\theta)}{\theta } + x \bigg| = \frac{N(-\tilde{b}\theta)}{\theta } -\tilde{b} \xrightarrow{p} 0,
 \end{equation}
 so putting everything together we have~\eqref{eq:unifconv}.

\subsection{A simple example with a two-cut phase} 
To conclude this section, we give the explicit angles defining the limit density and local limiting kernel for $H_{\theta\gamma}$ in a single parameter set-up. Let $\gamma_1 = 1$, let $\gamma_2 = \gamma$ and let $\gamma_r = 0$ for each $r >2$. This corresponds precisely to the ``probabilistic line'' of the time-evolving model studied by Bocini and Stéphan in~\cite{Bocini_Stephan_2020}, and the limit densities were already given in that paper. In this case the boundaries of the Fermi sea $I_x$ are simply given by solutions to 
\begin{equation}
D(\chi) - x= 8\gamma \cos^2 \chi + 2\cos \chi - 4\gamma -x = 0.
\end{equation}
Finding them amounts to finding the roots of $y \mapsto 8\gamma y^2 + 2y - 4\gamma -x$ between $-1$ and 1; the solution with discriminant $0$ lies at $\pm1$ for $\gamma = \mp \frac{1}{8}$, which distinguishes three regimes as follows (see also graphs in Figure~\ref{fig:4casesDetc}):
\begin{itemize}
\item \emph{Single Fermi sea:} For $ -\frac{1}{8}\leq \gamma \leq \frac{1}{8}$, there is a one-cut Fermi sea throughout the bulk. For $\gamma \neq 0$, the Fermi seas are given by
\begin{equation}
I_x = \begin{cases}
[-\pi,\pi], \qquad &x < 4\gamma -2\\
[-\chi_2,\chi_2] \quad \text{where} \quad \chi_2 = \arccos \bigg( \frac{-1+ \sqrt{1 + 8 x \gamma + 32 \gamma^2}}{8\gamma}\bigg) , \quad & 4\gamma -2 \leq x \leq 4 \gamma+2\\
\emptyset, \qquad &x > 4 \gamma +2
\end{cases}
\end{equation}
and for $\gamma = 0$ the result holds with $\chi_2 = \arccos \frac{x}{2}$ in the bulk (we implicitly have $\chi_1 = 0$).
\item \emph{Split Fermi sea at the right edge:} For $\gamma < -\frac{1}{8}$, the Fermi sea has two cuts before the right edge. It is given by
\begin{equation}
\varrho(x) = \begin{cases}
[-\pi,\pi], \qquad &x <  4\gamma -2\\
[-\chi_2,\chi_2] \quad \text{where} \: \chi_2 = \arccos \bigg( \frac{-1+ \sqrt{1 + 8 x \gamma + 32 \gamma^2}}{8\gamma}\bigg), \quad & 4\gamma -2\leq x \leq 4 \gamma + 2\\
[-\chi_2,-\chi_1] \cup [\chi_1,\chi_2] \\
\qquad \qquad \quad \text{where} \: \chi_{\substack{2\\1}} = \arccos \bigg( \frac{-1\pm \sqrt{1 + 8 x \gamma + 32 \gamma^2}}{8\gamma}\bigg) , \quad & 4 \gamma +2 \leq x \leq \frac{-1 - 32 \gamma^2}{8 \gamma}\\
\emptyset, \qquad &x >  \frac{-1 - 32 \gamma^2}{8 \gamma}
\end{cases}
\end{equation}
(we implicitly have $\chi_1 = 0$ in the second phase).
\item \emph{Split Fermi sea at the left edge:} 
For $\gamma > \frac{1}{8}$, there is a one-cut Fermi sea before the right edge and a two-cut one before the left edge. We have
\begin{equation}
I_x = \begin{cases}
[-\pi,\pi], \qquad &x < \frac{-1 - 32 \gamma^2}{8 \gamma} \\
[-\pi,-\chi_3]\cup [-\chi_2, \chi_2] \cup [\chi_3,\pi] \\
\qquad \qquad \quad \text{where} \: \chi_{\substack{3\\2}}=\arccos \bigg( \frac{-1\mp \sqrt{1 + 8 x \gamma + 32 \gamma^2}}{8\gamma}\bigg), \quad & \frac{-1 - 32 \gamma^2}{8 \gamma}  \leq x \leq 4 \gamma +2\\
[-\chi_2, \chi_2] \quad \text{where} \:\chi_2 = \arccos \bigg( \frac{-1+ \sqrt{1 + 8 x \gamma + 32 \gamma^2}}{8\gamma}\bigg) , \quad &  4\gamma -2\leq x \leq 4 \gamma+2\\
\emptyset, \qquad &x >  4 \gamma +2
\end{cases}
\end{equation}
(here we implicitly have $\chi_4 = \pi$ in the second phase, as well as $\chi_1 = 0$ throughout the bulk).
\end{itemize}

Let us mention how the resulting limit densities $\varrho$ go to $0$ at the right edge. Firstly, for $\gamma > -\frac{1}{8}$ where there is a single Fermi sea before the right edge, we have 
\begin{equation}
\varrho(x) \sim \frac{1}{\pi\sqrt{1+8\gamma}}(2+4\gamma - x)^{\frac{1}{2}}, \qquad x \to (4\gamma+2)^-,
\end{equation}
and we can write a  continuous function in $\gamma$ to describe the edge vanishing. 
For the marginal case $\gamma  = -\frac{1}{8}$, at which a split appears in the Fermi sea before the right edge, that function becomes singular. Now, for the edge vanishing we have 
\begin{equation}
\varrho(x) \sim \frac{\sqrt{2}}{\pi}\bigg(\frac{3}{2} - x\bigg)^{\frac{1}{4}} \qquad x \to \frac{3}{2}^-,
\end{equation}
which is the edge behaviour associated with \emph{order $2$ multicriticality} (see~\cite[Theorem~2]{BBW_2023}; the model with $\gamma_1=1$ and $\gamma_2 = -\frac{1}{8}$ is the $m=2$ ``minimal multicritical measure'' in the language of that paper). For $\gamma < -\frac{1}{8}$, where there are two cuts in the Fermi sea before the right edge, we have 
\begin{equation}
\varrho(x) \sim \frac{2}{\pi} \sqrt{\frac{8\gamma}{1-64\gamma^2}}\bigg(\frac{-1 - 32 \gamma^2}{8 \gamma} - x\bigg)^{\frac{1}{2}}, \qquad x \to \bigg(\frac{-1 - 32 \gamma^2}{8 \gamma}\bigg)^-.
\end{equation}
We recognise the same exponent as in the single Fermi sea case for $\gamma > -\frac{1}{8}$ (corresponding to $m=1$ non-multicritical behaviour), although the edge and scaling coefficients are different. We can find analogous results for how $1-\varrho(x)$ vanishes when $x \to -\tilde{b}^+$, with an exponent of $\frac{1}{4}$ at $\gamma = \frac{1}{8}$ only, and coefficients from the two-cut case for $\gamma > \frac{1}{8}$.

\section{Edge asymptotics for two-cut Fermi seas} \label{sec:edge}

\subsection{The edge of the limit density and heuristics for fluctuations} \label{sec:edgeheuristic}
Now we turn our attention to behaviour of fermions near the interface between the bulk and the empty region, where $\varrho \to 0$. This is at $x=b:= \max D$; since $D$ is a combination of cosines, its maximum is reached at some angles $\chi_b$ such that $D'(\chi_b) = 0$ and the first non-zero derivative of $D$ is of an even order and is negative. The number of $\chi_b$ for which $D$ is maximal is the number $n$ of cuts in $I_{b^-}$, the Fermi sea immediately before the right edge. For $n=1$, we have either $b = D(0)$ or $b = D(\pi)$. The edge behaviour in these cases was described in~\cite{BBW_2023} (although that paper uses the hypothesis that $I_x$ has one cut for all $x$, the proofs for the edge behaviour only use the weaker assumption that $I_{b^-}$ has one cut). 

Here we consider the next simplest case, where $n=2$. Then, there is a unique $\chi_b$ in $(0,\pi)$ such that 
\begin{equation}  \label{eq:Dbcond}
D(\chi_b) = D(-\chi_b) = \sum_{r\geq 1} 2 r \gamma_r \cos r \chi_b = b,
\end{equation}
or the maximum is reached at $0$ and $\pi$. We consider only the $\chi_b \in (0,\pi)$ case for simplicity. Throughout the following, we let $m$ be the positive integer satisfying 
\begin{equation} \label{eq:Ddcond}
D^{(p)}(\chi_b) = 0, \: \text{for} \: p = 1, \ldots, 2m-1, \qquad D^{(2m)}(\chi_b) = \sum_{r \geq 1} 2 (-1)^{m}r^{2m+1} \gamma_r \cos r \chi_b < 0
\end{equation}
and we fix the positive constant $d = - D^{(2m)}(\chi_b)/(2m)!$.

\paragraph*{Vanishing and fluctuation exponents} At the right edge of the bulk in our setup, the limit density is determined by  $\chi_2 - \chi_1$, where $\pi > \chi_2 >\chi_1 >0$ and each satisfy $D(\chi_i) = x$. As $x$ tends to $b$ from below, both $\chi_2,\chi_1$ tend to $\chi_b$. Developing $D(\chi_i)$ in $\chi_i$ near $\chi_b$ recovers
\begin{equation}
D(\chi_b) + \frac{D^{(2m)}(\chi_b)}{(2m)!} (\chi_i - \chi_b)^{2m} + O\big((\chi_i - \chi_b)^{2m+1}\big)= x, \qquad x<b.
\end{equation}
Since $\chi_2 > \chi_b$ and $\chi_1<\chi_b$, we find
\begin{equation}
 \varrho(x) = \frac{\chi_2 - \chi_b}{\pi} + \frac{\chi_b - \chi_1}{\pi} \sim \frac{2}{\pi} \bigg(\frac{b-x}{d}\bigg)^{\frac{1}{2m}}.
 \end{equation} 

 From this, we can estimate the scale of the fluctuations in the position of the rightmost occupied site $k_{\max}$ with the following heuristic. Let $k_{N}$ denote the position of the $N$th occupied site from the right (so that $k_{\max} = k_1$). Then if $N\sim u\theta$ for some finite $u$ we have $k_N \sim x\theta$ where $x$ is implicitly given by $u = \int_{x}^\infty \varrho(x')dx'$. As $x \to b^-$, we can estimate the integral of $\varrho$ by its edge behaviour, and we have that  $u \propto (b-x)^{\frac{2m+1}{2m}}$. Upon inverting this, we see that $
 k_N - b \theta \propto \theta u^{\frac{2m}{2m+1}} $
in this regime. Now, let us assume (without justification) that this estimate can be extended from finite $u$ to $u$ which scales as $\theta^{-1}$, so that $N \sim 1$. Inserting this into the above formula implies that
\begin{equation} \label{eq:scaleest}
k_{\max} - b\theta \propto \theta^{\frac{1}{2m+1}}
\end{equation}
as $\theta\to\infty$.

Let us note that the heuristic recalled at~\eqref{eq:edgeheuristic} for the edge scaling limiting is not physically meaningful when $I_{b^-}$ has two cuts and does not include $0$. Although we can make precisely the same arguments by letting the lattice coordinates scale in exactly the same way, with $k \sim \hat{b}\theta + x(\hat{d} \theta)^{\frac{1}{2\hat{m}+1}} $ where $\hat{m}$ is the smallest integer such that $\sum_{r} r^{2\hat{m} +1}\gamma_r \neq 0$  and $\hat{b} = 2\sum_r r\gamma_r$ and $  \hat{d} = \frac{2(-1)^{\hat{m}+1}}{(2\hat{m})!}\sum_r r^{2\hat{m} + 1}\gamma_r$; this gives the same continuous estimate for for $(d\theta)^{-\frac{1}{2\hat{m}+1}}H_{\theta\gamma}$ as $\theta \to \infty$. But now, $\hat{b}$ is not the edge of the limit density but rather a point in the bulk. Following the analysis in Section~\ref{sec:lsfromk}, the kernel $K_{\theta}$ has a non-trivial local limit  around $\hat{b} \theta$ given by an extended discrete sine kernel, so the implication that $(d\theta)^{\frac{1}{2\hat{m}+1}}K_{\theta}$ converges to $\mathcal{A}_{2\hat{m}+1}$ is false.  

\subsection{Oscillating edge limit} \label{sec:edgekernel}
Let us now consider the edge scaling limit of the kernel more formally, starting from the exact integral expression~\eqref{eq:kexact}. Here we follow precisely the steps of~\cite[Section~3.3]{BBW_2023}. In Section~\ref{sec:lsfromk}, we showed that if $k,\ell \sim b\theta$, $K_{\theta\gamma}(k,\ell)$ vanishes as $\theta \to \infty$. As there is an order $2m$ saddle point of $D$ at this point, we expect the leading order term to scale with $\theta^{-\frac{1}{2m+1}}$ from previous arguments. 

\begin{figure}

{\def\svgwidth{0.9\textwidth} 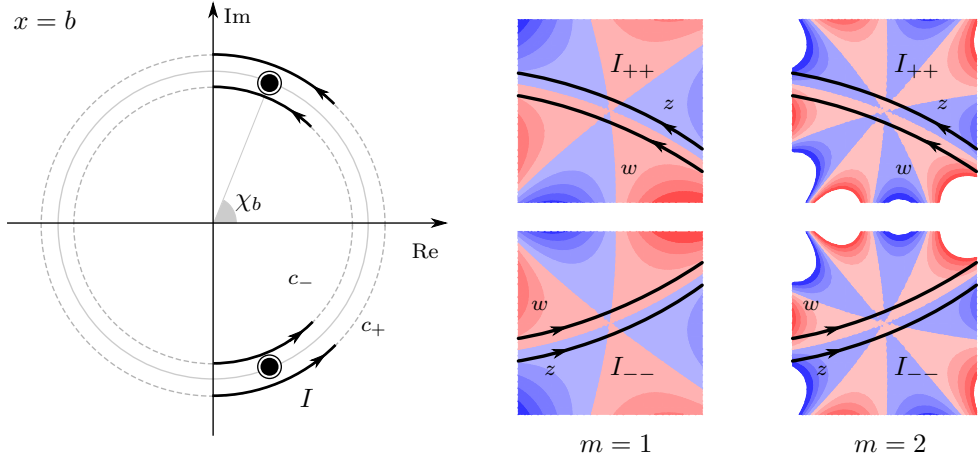 }
\caption{Left, the order $2$ saddle points of the action $S(z;b)$ for $\gamma_1 = 1$, $\gamma_2 = -\tfrac13$ and $\gamma_r = 0$ for $r>2$, at the right edge $b = \frac{41}{24}$. The central region $I$ of the integration contours $c_+\times c_-$ is  shown in black. In the centre, plots of the real part of $S(z;x)$ for $z$ near the saddle points $e^{i\chi_b}$ above and $e^{-i\chi_b}$ below with $\chi_b = \arccos \frac{3}{8}$, with blues where $\Re(S(z;b)) <0$ and reds where  $\Re(S(z;b)) >0$.  Right, the corresponding plots in the case where $\gamma_1,\gamma_2, \gamma_3,\gamma_4$ are tuned so that the action $S(z;b)$ has  order $4$ saddle points at $e^{\pm i \arccos \frac{3}{8} }$ on the right edge.} \label{fig:rightedge}
\end{figure}

We consider a regime at the same fluctuation scale predicted heuristically at~\eqref{eq:scaleest}, where $k \sim b\theta + x (d\theta)^{\frac{1}{2m+1}}$ and $\ell  \sim b\theta + y (d\theta)^{\frac{1}{2m+1}}$, for finite $x,y$ (note that although we reuse $x$ as a parameter here, it plays a very different role in this asymptotic regime to the $x$ of Section~\ref{sec:bulk}). In terms of the action defined at~\eqref{eq:action}, the kernel is
\begin{equation}
K_{\theta\gamma}(k,\ell) = \frac{1}{(2\pi i)^2} \oiint_{c_+,c_-} \frac{\exp[\theta S(z;b) - x(d\theta)^{\frac{1}{2m+1}} \log z]}{\exp[\theta S(w;b) - y(d\theta)^{\frac{1}{2m+1}} \log w]} [1+o(1)]\frac{\dd z \dd w}{z-w}
\end{equation}
as $\theta \to \infty$, where the $o(1)$ accounts for the difference between the continuous coordinates and their half-integer parts and the integrals in $z$ and $w$ are taken over contours $c_+$ passing just outside the unit circle and $c_-$ passing just inside it respectively. The action has saddle points on the unit circle at $e^{\pm i\chi_b}$, but we choose contours that only approach these points as $\theta \to \infty$, with 
\begin{equation}
c_\pm = \big\{ \exp\big[\pm (d\theta)^{-\frac{1}{2m+1}} + i\phi\big], \phi \in [-\pi,\pi]\big\}. 
\end{equation}
Along these contours, we have
\begin{equation}
\Re \big( S\big( \exp\big[\pm (d\theta)^{-\frac{1}{2m+1}} + i\phi\big]; b\big)\big) = \pm (D(\phi) - b)(d\theta)^{-\frac{1}{2m+1}} + O\big(\theta^{-\frac{3}{2m+1}}\big)
\end{equation}
uniformly in $\phi$. Since $D(\phi) < b$, for $\phi$ not equal to $\pm\chi_b$, the integrand is exponentially suppressed away from those points. Let us fix $\epsilon\in (0,\frac{2}{(2m+1)(2m+3)})$ and define the \emph{central region} of the integration domain as
\begin{equation}
I = \{(z,w) \in c_+ \times c_- :  \chi_b - (d\theta)^{\frac{1}{2m+1}} \leq|\arg z|, |\arg w| \leq \chi_b + (d\theta)^{\frac{1}{2m+1}}\}.
\end{equation}
It is the union of four disjoint domains $I_{++}, I_{--},  I_{+-}$ and $ I_{-+}$, which we label by
\begin{equation}
I_{\pm,\pm} = \{(z,w) \in I : \pm \arg z >0, \pm \arg w > 0\}.
\end{equation}
See Figure~\ref{fig:rightedge} for an illustration of the contours, their central region, and the saddle points.

First let us find the contribution from the complementary region $I^c = (c_+ \times c_-) \setminus I$. Since $D$ is maximal at $\pm \chi_b$, where it has order $2m$ saddle points, we can bound $D(\phi) -b$ above by $- C (\chi_b - |\phi|)^{2m} $ for some positive constant $C$.
Hence, for all $(z,w)\in I^c$ we see that
\begin{equation*}
e^{\theta \Re [S(z;b) - S(w;b)]}  = O(e^{-\theta^{2m\epsilon}/C}).
\end{equation*}
Since $I^c$ is a bounded domain and $1/(z-w) = O(\theta^{\frac{1}{2m+1}})$, the integral over $I^c$ decays  to $0$ exponentially fast as $\theta \to \infty$. 

It remains to find the integral over the central region $I$. 
Consider the integral over $I_{++}$, under the change of variables
\begin{equation} \label{eq:covariables}
z = \exp\big[i\chi_b + \zeta (d\theta)^{-\frac{1}{2m+1}}\big], \quad w = \exp\big[i\chi_b + \omega (d\theta)^{-\frac{1}{2m+1}}\big]
\end{equation}
In terms of $I_\theta = [-(d\theta)^\epsilon,(d\theta)^\epsilon]$, the domain of integration in $\zeta$ is $i I_\theta + 1$ and in $\omega$ it is $iI_\theta -1$. Taylor expanding the action of $z$ in $I_{++}$, we have 
\begin{equation}
\begin{split}
S\big(\exp\big[i\chi_b +\zeta (d\theta)^{-\frac{1}{2m+1}}\big];b\big) = &S(e^{i\chi_b};b)+ D(\chi_b) \zeta (d\theta)^{-\frac{1}{2m+1}} - i D'(\chi_b) \zeta (d\theta)^{-\frac{1}{2m+1}} + \ldots \\
&  + (-1)^m \frac{D^{(2m)}(\chi_b)}{2m+1} \zeta^{2m+1} (d\theta)^{-1} - b\zeta(d\theta)^{-\frac{1}{2m+1}} + o(\theta^{-1}),
\end{split}
\end{equation}
where the error is more precisely $O(\theta^{(2m+3)\epsilon - \frac{2m+3}{2m+1}})$, uniformly in $\zeta$. Recalling the conditions~\eqref{eq:Dbcond} and~\eqref{eq:Ddcond} satisfied by $D$, the dominant term on the right hand side simplifies to $S(e^{i\chi_b};b)  + \frac{(-1)^m }{2m+1}\frac{\zeta^{2m+1}}{\theta}$. This is precisely what we obtain from the action at the $z=1 $ saddle point in the case where $I_{b^-}$ has one cut. 
Proceeding similarly for the action of $w$, noting that $x\log z = x(i\chi_b + \zeta (d\theta)^{-\frac{1}{2m+1}})$
 and that 
\begin{equation*}
\frac{\dd z \dd w}{ z- w} = e^{i\chi_b}(d\theta)^{-\frac{1}{2m+1}}\frac{\dd \zeta \dd \omega}{\zeta -\omega} + O(\theta^{-\frac{2}{2m+1}}),
\end{equation*}
the integral over $I_{++}$ is 
\begin{equation*}
 (d\theta)^{-\frac{1}{2m+1}} \exp[i\chi_b (d\theta)^{\frac{1}{2m+1}}(x-y)+i\chi_b]\int\displaylimits_{iI_\theta - 1} \int\displaylimits_{iI_\theta +1} \frac{\exp\big[ \frac{(-1)^m }{2m+1}\zeta^{2m+1} - x \zeta\big] }{\exp\big[ \frac{(-1)^m }{2m+1}\omega^{2m+1} - y \omega\big]}[1+o(1)] \frac{\dd \zeta \dd \omega}{\zeta -\omega} 
\end{equation*}
as $\theta \to \infty$. The prefactor is new to the case where $I_{b^-}$ has two cuts and $\chi_b$ is non-zero. The dominant term of the double integral is equivalent to the order-$m$ Airy function in the limit as $\theta\to \infty$ and $I_\theta \to \R$: upon inserting $1/(\zeta-\omega) = \int_0^\infty e^{v(\zeta-\omega)}dv$, we see that
\begin{equation}
 \int\displaylimits_{i\R - 1} \int\displaylimits_{i\R +1} \frac{\exp\big[ \frac{(-1)^m }{2m+1}\zeta^{2m+1} - x \zeta\big] }{\exp\big[ \frac{(-1)^m }{2m+1}\omega^{2m+1} - y \omega\big]}\frac{\dd \zeta \dd \omega}{\zeta -\omega} = \int_0^\infty \Ai_{2m+1}(x-v)\Ai_{2m+1}(x-v) \dd v.
\end{equation}

 For the integral over $I_{--}$, we make the similar change of variables 
\begin{equation*} 
z = \exp\big[-i\chi_b + \zeta (d\theta)^{-\frac{1}{2m+1}}\big], \quad w = \exp\big[-i\chi_b + \omega (d\theta)^{-\frac{1}{2m+1}}\big]
\end{equation*}
and find the same double integral with a prefactor of $(d\theta)^{\frac{1}{2m+1}}\exp[-i\chi_b (d\theta)^{\frac{1}{2m+1}}(x-y)]$. For the integrals over $I_{+-}$ and $I_{-+}$ respectively, the changes of variables
\begin{equation*} 
z = \exp\big[\pm i\chi_b + \zeta (d\theta)^{-\frac{1}{2m+1}}\big], \quad w = \exp\big[\mp i\chi_b + \omega (d\theta)^{-\frac{1}{2m+1}}\big]
\end{equation*}
give
\begin{equation*}
\frac{\dd z \dd w}{ z- w} = (d\theta)^{-\frac{2}{2m+1}}\frac{\dd \zeta \dd \omega}{2 i \sin \pm \chi_b} + O(\theta^{-\frac{3}{2m+1}});
\end{equation*}
we see that the contributions from $I_{+-}$ and $I_{-+}$ are each $O(\theta^{-\frac{2}{2m+1}})$. 
Putting together the leading order contributions in $\theta$, from $I_{++}$ and $I_{--}$ only, we have
\begin{equation}
 (d\theta)^{\frac{1}{2m+1}}K_{\theta\gamma}( k, \ell) =  2\cos \chi_b (d\theta)^{\frac{1}{2m+1}}(x-y) \int\displaylimits_{iI_\theta - 1} \int\displaylimits_{iI_\theta +1} \frac{\exp\big[ \frac{(-1)^m }{2m+1}\zeta^{2m+1} - x \zeta\big] }{\exp\big[ \frac{(-1)^m }{2m+1}\omega^{2m+1} - y \omega\big]} [1+o(1)]\frac{\dd \zeta \dd \omega}{\zeta -\omega} 
\end{equation}
as $\theta\to\infty$.
This partially resembles the integral obtained in the case where the Fermi sea $I_{b^-}$ has one cut. We recognise the overall scaling factor of $(d\theta)^{-\frac{1}{2m+1}}$, and the same double integral which converges to $\mathcal{A}_{2m+1}$, up to an overall factor of $2$.
However, we also have a rapidly oscillating term, which does \emph{not} have any limit as $\theta \to \infty$. Unlike in the single Fermi sea case, we cannot specify a limiting determinantal point process in the edge scaling regime.

\paragraph*{Towards the general case} Let us mention the cases we have not treated here. In the two-cut case, we did not treat the situation where $D$ is maximised at $0$ and $\pi$, but it is straightforward to see that there is no oscillating term in this case; rather, we have $(d\theta)^{\frac{1}{2m+1}}K_{\theta\gamma}( k, \ell) \to 2\mathcal{A}_{2m+1}(x,y)$ as $\theta\to\infty$. This case is in some sense more similar to one-cut case, as the heuristic~\eqref{eq:edgeheuristic} still makes sense physically.  
We also did not consider the case where $I_{b^-}$ has more than two cuts. Then, the maximisers of $D$ may be at saddle points of different order, and some may be at $0$ or $\pi$. It is straightforward to see that only the lowest order saddle points will determine the edge scaling regime and corresponding limit of the kernel, and that any pair of complex conjugate saddle points away from $0$ or $\pi$ will add a prefactor of the form $2\cos \chi_b (d\theta)^{1/2m+1}$ to the prefactor, and any saddle point at $0$ or $\pi$ will just add  $1$.

\subsection{A limiting distribution} \label{sec:edgefluctuation}

Let us turn our attention to the asymptotic fluctuations in $k_{\max}$ around $b$ in the two-cut Fermi sea case. By the inclusion-exclusion principle, the cumulative distribution of $k_{\max}$ may be written as the Fredholm determinant 
\begin{equation}
\begin{split}
\P_{\theta\gamma}(k_{\max} < \ell) = \det(1 - K_{\theta\gamma})_{l^2(\ell+\frac12,\ell+\frac32, \ldots)} = \sum_{n=0}\frac{(-1)^n}{n!}  \sum_{k_1 = \ell + \frac{1}{2}}^\infty \cdots  \sum_{k_n = \ell + \frac{1}{2}}^\infty \det_{1 \leq i,j \leq n} K_{\theta\gamma}(k_i,k_j)
\end{split}
\end{equation}
(in terms of an integer $\ell$). We are interested in the limit of this  distribution as $\theta \to \infty$ in a regime where $\ell  \sim b\theta + s(d\theta)^{\frac{1}{2m+1}}$. Although we do not have a true limit for the kernel, we note that it only appears integrated over in the Fredholm determinant. Informally, we may estimate the limiting distribution by averaging over the rapidly oscillating terms.

We start from the logarithm of the above Fredholm determinant, which is 
\begin{equation}
\log \P_{\theta\gamma}(k_{\max} < \ell) = - \sum_{n = 2}^\infty \frac{1}{n} \sum_{k_1 = \ell + \frac{1}{2}}^\infty \hspace*{-0.5em}\cdots  \hspace*{-0.5em} \sum_{k_n = \ell + \frac{1}{2}}^\infty  \hspace*{-0.5em}K_{\theta\gamma}(k_1,k_2) K_{\theta\gamma}(k_2,k_3) \cdots K_{\theta\gamma}(k_n,k_1).
\end{equation}
Here, we see the diagonal term contributes negligibly to all but the $n=1$ term. For that term, we have
\begin{equation}
\sum_{k = \ell +\frac12}^\infty K_{\theta\gamma}(k,k)  \to 2\int_{s}^\infty \mathcal{A}_{2m+1}(x,x) \dd x 
\end{equation}
(the factor of $(d\theta)^{-\frac{1}{2m+1}}$ from the limiting kernel compensates the factor of $(d\theta)^{\frac{1}{2m+1}}$ from the differential, and we have dominated convergence from an exponential bound on the kernel). 
For each of the $n \geq 2$ terms, let us suppose that we can factor out the oscillating term from the kernel; because that term varies much more rapidly than the rest of the kernel, we estimate that we have
\begin{equation}
 \begin{split}
 \sum_{k_1 = \ell+ \frac{1}{2}}^\infty \hspace*{-0.5em}\cdots  \hspace*{-0.5em} &\sum_{k_n = \ell + \frac{1}{2}}^\infty  \hspace*{-0.5em}K_{\theta\gamma}(k_1,k_2) K_{\theta\gamma}(k_2,k_3) \cdots K_{\theta\gamma}(k_n,k_1) \\
 &\to C_n 2^n \int_{s}^\infty \hspace*{-0.5em}\cdots  \int_{s}^\infty \hspace*{-0.5em} \mathcal{A}_{2m+1}(x_1,x_2) \mathcal{A}_{2m+1}(x_3,x_4) \cdots \mathcal{A}_{2m+1}(x_n,x_1)\dd x_1 \cdots \dd x_{n} 
  \end{split}
 \end{equation} 
 as $\theta\to \infty$, where $C_n$ accounts for the averaging over the product of $n$ oscillating terms. Averaging the oscillating term over a period of $2\pi/\chi_b$, we have
 \begin{equation}
 \begin{split}
 C_n = \bigg(\frac{\chi_b}{2\pi}\bigg)^{n} \hspace*{-0.5em} \int_0^{\frac{2\pi}{\chi_b}} \hspace*{-0.7em}\cdots \hspace*{-0.2em} \int_0^{\frac{2\pi}{\chi_b}} \hspace*{-0.7em} \cos \chi_b(x_1 - x_2 ) \cos \chi_b(x_2 - x_3 ) \cdots \cos \chi_b(x_n - x_1 ) \dd x_1 \cdots \dd x_{n} = \frac{1}{2^{n-1}}.
 \end{split}
 \end{equation}
With this, we argue that we have
\begin{equation}
\log \P_{\theta\gamma}(k_{\max} < \ell) \to -2\sum_{n=1}^{\infty}  \int_{s}^\infty \hspace*{-0.5em}\cdots  \int_{s}^\infty \hspace*{-0.5em} \mathcal{A}_{2m+1}(x_1,x_2) \cdots \mathcal{A}_{2m+1}(x_n,x_1) \dd x_1 \cdots \dd x_{n}
\end{equation}
as $\theta \to \infty$. We recognise the first term as the log of a continuous Fredholm determinant. Taking the exponential, we estimate the limiting distribution as
\begin{equation}
\lim_{\theta \to \infty} \P_{\theta\gamma} \bigg( \frac{k_{\max} - b\theta}{(d\theta)^{\frac{1}{2m+1}}}<s \bigg) =  \det (1 -  \mathcal{A}_{2m+1})^2_{L^2([s,\infty))} = F_{2m+1}^2(s). 
\end{equation}
We conclude that the limiting distribution of $k_{\max}$ is the same as the maximum element of two independent copies of the determinantal point process obtained in the edge scaling limit from the ground state of a $H_{\theta\gamma}$ for which $I_{b^-}$ has one cut and the limit density has the same vanishing exponent of $\frac{1}{2m}$. It is natural to conjecture that the analogous limiting process is an $\alpha$-\emph{determinantal point process} at $\alpha = -\frac12$, see~\cite{Cunden_2019} and references therein. Indeed, the factors appearing from averaging over oscillations are consistent with such a description.

The case where $I_{b^-}$ has more than $2$ cuts is more subtle as the contributing saddle points may be of different order, but we can generally expect the oscillating factor to be sum of $n$ cosines of the form above, with different angles $\chi_b$. Averaging over such a term gives $C_n = n/2^{n-1}$; we conjecture that we obtain the limiting distribution $F_{2m+1}^n(s)$ when $I_{b^-}$ has $n$ cuts.

\section{Unitary matrix models with multi-cut potentials} \label{sec:matrixmodels}

\subsection{From edge distributions to unitary matrix models} \label{sec:matrixintegraldists}
In this section we revisit the edge distribution $\P_{\theta\gamma}(k_{\max} < \ell)$ from another perspective. By a formula of Borodin and Okounkov~\cite{Borodin_Okounkov_2000}, this is proportional to an $\ell \times \ell$ Toeplitz determinant, with
\begin{equation} \label{eq:boformula}
e^{\theta \sum_r r \gamma_r^2}\P_{\theta\gamma}(k_{\max} < \ell) = \det_{1 \leq i,j \leq \ell} f_{j-i}
\end{equation}
where the  matrix entries are given by the generating function
\begin{equation}
\sum_{n \in \Z} f_{n} z^n = e^{-\theta \sum_{r} (-1)^{r} \gamma_r ( z^r + z^{-r })}
\end{equation}
(see Appendix~\ref{app:schur} for a short proof). 
As shown for instance in~\cite{Baik_Rains_2001}, this Toeplitz determinant can be written as an integral over the $\ell \times \ell$ unitary matrices, as follows. The entries $f_n$ may be extracted from their generating function using contour integrals; we may take them over the unit circle $c_1$, and we have
\begin{equation}
e^{\theta \sum_r r \gamma_r^2}\P_{\theta\gamma}(k_{\max} < \ell) =  \det_{1 \leq i,j \leq \ell } \frac{1}{(2\pi i)} \oint\displaylimits_{c_1} u^{i-j} e^{-\theta \sum_{r} (-1)^{r} \gamma_r ( u^r + u^{-r })} \frac{\dd u}{u}. 
\end{equation}
By an application of the Andreïef identity (see e.g.~\cite{Forrester_2019}),  this can be rewritten as the $\ell$-fold integral on $c_1$
\begin{equation*}
\frac{1}{(2\pi i)^\ell \ell !}\oint\displaylimits_{c_1} \cdots  \oint\displaylimits_{c_1} \det_{1 \leq i,j \leq \ell } u_j^{i-1} e^{-\theta \sum_{r} (-1)^{r} \gamma_r u_j^r }\det_{1 \leq i,j \leq \ell } u_j^{i-1} e^{-\theta \sum_{r} (-1)^{r} \gamma_r  u_j^{-r }} \frac{\dd u_1}{u_1} \cdots \frac{\dd u_\ell}{u_\ell}.
\end{equation*}
We recognise two copies of the Vandermonde determinant 
\begin{equation}
\det_{1 \leq i,j \leq \ell } u_i^{j-1} = \prod_{i<j}(u_j - u_i)
\end{equation}in the integrand. Since each $u_i$ is on the unit circle and $u_i^{-1} = \bar{u}_i$, we see that the Toeplitz determinant may be written as
\begin{equation}\label{eq:weyl}
\begin{split}
 e^{\theta \sum_r r \gamma_r^2}\P_{\theta\gamma}(k_{\max} < \ell) &=\frac{1}{(2\pi i)^\ell \ell !} \oint_{c_1} \hspace{-0.3em}\cdots  \oint_{c_1} \prod_{i=1}^\ell e^{-\theta (-1)^{r} \gamma_r ( u_i^r + u_i^{-r })}  \prod_{i<j}|u_i - u_j|^2 \frac{du_1}{ u_1} \cdots\frac{du_\ell}{u_\ell} \\
 &=  \int_{\mathcal{U}(\ell)} e^{-\theta \tr \sum_r (-1)^{r} \gamma_r ( U^r + U^{-r })} \mathcal{D} U 
 \end{split}
\end{equation}
where for the second equality, we recognise that by Weyl's integration formula (see e.g.~\cite[Chapter~1]{Meckes_2019}) this is an integral over the unitary group $\mathcal{U}(\ell)$ with respect to the Haar measure $\mathcal{D}U$ (this formula amounts to a change of variables from the eigenvalues $u_1, \ldots,u_\ell$ of a unitary matrix ${U}$).

From this expression, we can interpret the distribution $Z_\ell := e^{\theta \sum_r r \gamma_r^2}\P_{\theta\gamma}(k_{\max} < \ell)$ as the partition function of the unitary matrix ensemble with probability density
\begin{equation}
p_{\theta\gamma;\ell}^m(U) = \frac{1}{Z_\ell}e^{-\theta \tr \sum_r (-1)^{r} \gamma_r ( U^r + U^{-r })}.
\end{equation}
This is a generalisation of the Gross--Witten--Wadia model~\cite{Gross_Witten_1980,Wadia_1980}, and of the multicritical Periwal--Shevitz models~\cite{Periwal_Shevitz_1990,Periwal_Shevitz_1990_2}.

\subsection{Limiting eigenvalue densities} \label{sec:limiting_eigenvalue_densities}

Let us consider how the eigenvalues of an $\ell \times \ell$ random unitary matrix $U$ under the probability density $p_{\theta\gamma;\ell}(U)$ behave as $\ell\to\infty$ when the coupling is set to $\theta = \ell / x$ for positive $x$. For the following we let  $V(z):=\sum_r \gamma_r z^r$ denote the \emph{potential} of the model. We will work in terms of the arguments $\alpha_i$ of the eigenvalues $e^{i\alpha_i}$ of $U$, which we order as $-\pi \leq \alpha_1 \leq \alpha_2 \leq \ldots \leq \alpha_\ell \leq \pi$. From Weyl's formula~\eqref{eq:weyl},  the induced joint probability density on the $\alpha_i$ is 
\begin{equation}
 p_{\gamma/x;\ell}( \alpha_1,\ldots,\alpha_\ell) = \frac{4^{\ell(\ell+1)/2}}{Z_\ell (2\pi)^\ell \ell!}e^{-\frac{\ell}{x} \sum_{j=1}^{\ell} [V(-e^{i\alpha_j})+V(-e^{-i\alpha_j}) ]} \prod_{j<k} \big|\sin \frac{\alpha_j-\alpha_k}{2}\big|^2
\end{equation}
on the ordered arguments  of the eigenvalues $e^{i\alpha_j}$ of $U$ with respect to $ d\alpha_1 \cdots d\alpha_\ell$. 
Let us define the non-decreasing function
\begin{equation}
\alpha(u):= \alpha_{\lfloor u \ell\rfloor}
\end{equation} 
encoding the arguments of the random eigenvalues. As $\ell \to \infty$, we find a limiting eigenvalue density by optimising the functional 
\begin{equation}
-\frac{1}{x}\int_0^1 [ V(-e^{i\alpha(u)})+V(-e^{-i\alpha(u)})] du + \fint_{0}^1 \fint_{0}^1 \log \big|\sin \frac{\alpha(u)-\alpha(v)}{2}\big| du \, dv
\end{equation}
where  $\fint$ denotes the Cauchy principal part. This approach was introduced in~\cite{BIPZ_1978}, and adapted to a similar density on unitary matrices in~\cite{Gross_Witten_1980}; here, we adapt a computation appearing in~\cite[Section~4.2]{BBW_2023}.
The limit of the function $\alpha(u)$  is related to the limiting density by $\rho(\alpha) = du/d\alpha$. The saddle point equation for the functional above is
\begin{equation} \label{eq:equillibriumformula}
 \frac{i}{x} \big[e^{i\alpha}V'(-e^{i\alpha})-e^{-i\alpha}V'(-e^{-i\alpha})\big]= \fint_{0}^1 \cot \frac{\alpha-\alpha(v)}{2} dv = \fint_{-\beta_c}^{\beta_c} \rho(\beta) \cot \frac{\alpha-\beta}{2} d\beta 
\end{equation}
where the support $[-\beta_c,\beta_c]$ of $\rho$ is also to be determined. Inserting $e^{i (\alpha+\pi)}$ for $-e^{i\alpha}$ we have 
\begin{align}
\fint_{-\beta_c}^{\beta_c} \rho(\beta) \cot \frac{\alpha-\beta}{2} d\beta = -\frac{1}{x} \sum_{r\geq 1} 2 r \gamma_r \sin r (\alpha - \pi) . \label{eq:densitycond}
\end{align}

We will find $\rho(\alpha)$ in the ``supercritical'' phase, corresponding to the empty regime of the lattice fermion model (although it is feasible to find the limit in the ``subcritical'' regime as well; see the final equations of~\cite{Periwal_Shevitz_1990} for an explicit formula for the density and its support below criticality in any degree $4$ potential). Let $x$ be sufficiently large so that $\rho$ is supported on $[-\pi,\pi]$. Then, following the steps of~\cite[Page 449]{Gross_Witten_1980}, we may note that 
\begin{equation}
\fint_{-\pi}^{\pi} \cos(\beta-\pi) \cot \frac{\alpha-\beta}{2} d\beta = 2\pi \sin(\alpha-\pi)
\end{equation}
 and hence find the density
 \begin{equation} \label{eq:evdensityformula}
 \rho(\alpha) = \frac{1}{2 \pi} \bigg[ 1 - \frac{1}{x} \sum_{r\geq 1} 2 r \gamma_r \cos r (\alpha - \pi) \bigg] = \frac{1}{2 \pi} \bigg[1-\frac{1}{x}D(\alpha-\pi)\bigg].
 \end{equation}
For each maximiser $\chi_b$ of $D$, we see that $\rho(\chi_b + \pi)$ goes to $0$ as $x \to b^+$. So, the number of cuts in the Fermi sea $I_{b^-}$ determines the number of cuts appearing in the support of the limiting eigenvalue density at the phase transition. Moreover, at $x=b$, if $\chi_b$ is an order $2m$ saddle point of $D$, we immediately have that $\rho$ vanishes as 
\begin{equation}
\rho(\alpha)  \sim \frac{1}{2 \pi } \frac{d}{b}(\alpha-\pi-\chi_b)^{2m}, \qquad \alpha \to \pi + \chi_b. 
\end{equation}

\section{Conclusion and perspectives}

We have generalised a number of asymptotic results for lattice fermion models to the case where the Fermi seas of the model can have multiple cuts. In the bulk, this gives more interesting, non-analytic limit densities, and an extended discrete sine kernel in the local limit. At the edge, however, we have shown that there is no true limiting kernel, and that averaging over the oscillating kernel gives new asymptotic distributions for the rightmost fermion position. The resulting distribution corresponds to the distribution of the maximum element of independent copies of the Airy process or its order-$m$ analogue. The lattice fermion models with split Fermi seas are analogous to matrix models whose limiting eigenvalue densities have multi-cut support. On the one hand, the oscillation phenomenon we found for the kernel resembles one found in multi-cut Hermitian matrix models in~\cite{Bonnet_David_Eynard_2000}; on the other, these models are in exact correspondence with multi-cut unitary matrix models.

The new limit densities have internal interfaces where the Fermi sea splits, which could give rise to interesting new phenomena; although we know the leading order of the kernel, we have yet to investigate the fluctuations. Various questions remain open about the new edge fluctuations given by powers of the order-$m$ Tracy--Widom distribution. These distributions are related to Painlevé II equations, following~\cite{Cafasso_Claeys_Girotti_2019}; in the case of ``minimal multicritical models'' found in~\cite{BBW_2023}, the relevant Painlevé II equation has been derived by Chouteau and Tarricone~\cite{Chouteau_Tarricone_2022} from a (heuristic) continuum limit of a discrete recurrence relation shown by the same authors to be satisfied by any distribution $\P_{\theta\gamma}(k_{\max} < \ell)$. It would be interesting to find the corresponding continuum limit in the multi-cut case. Equally, it would be worth interpreting the distributions $F_{2m+1}^n$ in terms of the phase transition in the corresponding unitary matrix model; although we have the same critical exponents, we have a new prefactor in the free energy. 

\section*{Acknowledgements}
The author thanks Jérémie Bouttier for extensive guidance on this work, and thanks Dan Betea, Neil O'Connell, Giulio Ruzza, Jean-Marie Stéphan and Sofia Tarricone for helpful discussions and feedback. 
\appendix

\section{Random integer partition formulation} \label{app:schur}

\paragraph*{Schur function and Schur measure} In this appendix we relate our models to the Hermitian Schur measure $\P_{\theta\gamma}(\lambda)$,  introduced (in a more general setting) by Okounkov in~\cite{Okounkov_2001}. We begin with some definitions. Let $\lambda = (\lambda_1 \geq \lambda_2 \geq \ldots) $ be a partition, that is a weakly decreasing sequence of non-negative integers with a finite number $\ell(\lambda)$ of non-zero terms. For a given real sequence $t = (t_1,t_2,\ldots)$, the \emph{Schur function} $s_\lambda[t]$ may be defined (via the Jacobi--Trudi identity, see e.g.~\cite{Macdonald_1998}) as
\begin{equation}
s_{\lambda}[t] = \det_{1 \leq i,j \leq \ell(\lambda)} h_{\lambda_i - i+ j}[t]
\end{equation}
in terms of the functions $h_n$ generated by
\begin{equation}
\sum_{n \in \Z} h_n[t] z^n = e^{\sum_r t_r z^r}
\end{equation}
(we have $h_0 = 1$ and $h_n = 0$  for all $n <0$).
In the language of symmetric functions, the $h_n[t]$ are the complete homogeneous functions in the Miwa times $t_r$. If the sequence $t_r$ has finite support, by the Cauchy identity we have $\sum_\lambda s_\lambda [t]^2 = e^{\sum_r rt_r^2}$ where the sum is taken over all partitions. Then, we define the \emph{Hermitian Schur measure}
\begin{equation}
\P_{t}(\lambda) := e^{-\sum_r rt_r^2} s_{\lambda}[t]^2
\end{equation}
on all partitions. 

\paragraph*{Correspondence with lattice fermions} Let us relate the Schur measure to a measurement of every site in the ground state of $H_{\theta\gamma}$, and in particular prove~\eqref{eq:schurmeasS}. We let $\ket{\lambda}$ denote the state
\begin{equation}
\ket{\lambda} := c_{\lambda_{\ell(\lambda)} - \ell(\lambda) + \frac12}^{\dagger}c_{-\ell(\lambda) + \frac12} \ldots c_{\lambda_2 - \frac32}^{\dagger}c_{-\frac32}c_{\lambda_1 - \frac12}^{\dagger}c_{-\frac12}\ket{\emptyset}
\end{equation}
with the sites labelled $\{\lambda_i - i +\frac{1}{2}, i \in \Z_{>0}\}$ filled. Recalling the notation introduced at~\eqref{eq:unitaryop}, the probability of observing the ground state of 
$H_{\theta\gamma}$ in the state $\ket{\lambda}$ is $|\bra{\emptyset}\mathcal{U}_{\theta\gamma}^{-1} \ket{\lambda}|^2 $. To compute this term, we first perform a normal ordering of $\mathcal{U}_{\theta\gamma}$ by the Baker--Campbell--Hausdorff formula, which gives 
\begin{equation}
 \mathcal{U}_{\theta\gamma} = e^{\theta \sum_r \gamma_r (a_r^\dagger -  a_r)} = e^{-\frac{1}{2}\theta^2 \sum_r r \gamma_r^2}e^{\theta \sum_r \gamma_r a_r^\dagger} e^{-\theta \sum_r \gamma_r a_r} .
\end{equation}
Hence we have $\bra{\emptyset}\mathcal{U}_{\theta\gamma}^{-1} = e^{-\frac{1}{2}\theta^2 \sum_r r \gamma_r^2} \bra{\emptyset}e^{\theta \sum_r \gamma_r a_r}$. From the commutators~\eqref{eq:accomms} and the generating function of the $h_n[\theta\gamma]$, we obtain
\begin{equation}
e^{\theta \sum_r \gamma_r a_r} c^\dagger_k = \sum_{i \in \Z} h_i[\theta\gamma ]c^\dagger_{k-i} e^{\theta \sum_r \gamma_r a_r},  \qquad e^{\theta \sum_r \gamma_r a_r}  c_k = \sum_{i \in \Z} h_i[\theta\gamma]c_{k+i} e^{\theta \sum_r \gamma_r a_r} .
\end{equation}
In terms of the linear combinations of creation and annihilation operators $\hat{c}^\dagger_k := \sum_{i \in \Z} h_i[\theta\gamma ]c^\dagger_{k-i}$ and $c_k :=  \sum_{i \in \Z} h_i[\theta\gamma]c_{k+i}$, we have 
\begin{equation}
\begin{split}
\bra{\emptyset}\mathcal{U}_{\theta\gamma}^{-1} \ket{\lambda} &= e^{-\frac{1}{2} \theta^2 \sum_r r \gamma_r^2} \bra{\emptyset}  \hat{c}_{\lambda_{\ell(\lambda)} - \ell(\lambda) + \frac12}^{\dagger}\hat{c}_{-\ell(\lambda) + \frac12} \ldots \hat{c}_{\lambda_2 - \frac32}^{\dagger}\hat{c}_{-\frac32}\hat{c}_{\lambda_1 - \frac12}^{\dagger}\hat{c}_{-\frac12}\ket{\emptyset} \\
&= e^{-\frac{1}{2} \theta^2 \sum_r r \gamma_r^2}\det_{1 \leq i,j \leq \ell(\lambda)} \bra{\emptyset} \hat{c}_{\lambda_i - i + \frac12}^{\dagger}\hat{c}_{-j + \frac12}\ket{\emptyset}
\end{split}
\end{equation}
where the final equality is found by applying Wick's theorem (see e.g.~\cite{Gaudin_1960}). The matrix element is 
\begin{align}
\bra{\emptyset}  \hat{c}^\dagger_{\lambda_i  - i  + \frac{1}{2}} \hat{c}_{-j + \frac{1}{2}} \ket{\emptyset} &= \sum_{m, n \in \Z} h_m[\theta\gamma] h_n[\theta\gamma] \delta_{\lambda_i  - i  -m,n-j } \notag \\
&= \sum_{n} h_{\lambda_i  - i + j - n}[\theta\gamma] h_n[\theta\gamma] = h_{\lambda_i  - i + j}[\theta\gamma]
\end{align}
(the final equality is a generic a property of the complete homogeneous functions). We recognise that the determinant of this element is a Schur function, and obtain
\begin{equation}
|\bra{\emptyset} \mathcal{U}_{\theta\gamma}^{-1} \ket{\lambda}|^2 = e^{-\theta^2 \sum_r r \gamma_r^2} s_\lambda[\theta\gamma].
\end{equation}
This gives us precisely~\eqref{eq:schurmeasS}. 

\paragraph*{Limit shape of  a random partition} Let us describe the large scale behaviour of a random partition under $\P_{\theta\gamma}$ as $\theta \to \infty$. The \emph{Young diagram} of a partition $\lambda$ is an arrangement of $|\lambda| = \sum_i \lambda_i$ boxes in left-aligned rows with $\lambda_i$ boxes in the $i$th row, counted from the bottom up (in the French convention). The \emph{rescaled profile} $\psi_{\lambda,\theta}$ of a partition $\lambda$ is defined from the implicit relations
\begin{equation} \label{eq:profilephis}
v = v(u) := \tfrac{1}{\theta} \lambda_{\lfloor  \theta u \rfloor+1}, \; u \in (0, \infty) \quad\text{and}\quad u = u(v) := \tfrac{1}{\theta} \lambda'_{\lfloor  \theta v \rfloor+1}, \;  v \in (0, \infty)
\end{equation}
(with $\lfloor \cdot \rfloor$ denoting the floor function) through a change of coordinates 
\begin{equation} \label{eq:changetohookcoords}
\psi_{\lambda,\theta}(x) = u+v, \qquad x = v-u;
\end{equation}
It traces the upper edge of the Young diagram of $\lambda$ rotated anticlockwise by $45^\circ$ and centred on the origin, in coordinates scaling with $1/\theta$.  

Let $\lambda$ be a random partition under $\P_{\theta\gamma}$, and let $N(k)$ denote the random number of sites greater than $k$ observed to be occupied in a measurement of the ground state of $H_{\theta\gamma}$. Then, $\psi_{\lambda;\theta}(x) $ is equivalent in law  to $ x +{2} N(x\theta)/{\theta}.$ From the convergence in probability of $N(x\theta)/\theta$, we have that $ \psi_{\lambda;\theta}(x)$ converges uniformly in probability to 
 \begin{equation}
  \Omega(x) = x + 2 \int_x^\infty \varrho(x') \dd x'
 \end{equation}
 as $\theta \to \infty$, where $\varrho$ is the limit density given at~\eqref{eq:ls}. This may be formulated as a theorem, generalising~\cite[Theorem~2]{BBW_2023}: the proof consists of the rigorous saddle point analysis shown in Section~\ref{sec:lsfromk} and an application of~\cite[Lemma~11]{BBW_2023}. 

\paragraph*{Edge statistics of a random partition} Finally, let us consider how the first part $\lambda_1$ of a random partition $\lambda$ behaves under $\P_{\theta\gamma}$. We immediately have that $\lambda_1$ is equivalent in law to $k_{\max} + \tfrac{1}{2}$. Then, we can repeat the arguments of Section~\ref{sec:edge} and claim  that if $I_{b^-}$, the Fermi sea immediately left of $b = \max D$, has two cuts and the limit shape behaves as 
\begin{equation}
\Omega'(x) \sim - \frac{4}{\pi} \bigg( \frac{b-x}{d}\bigg)^{\frac{1}{2m}} \qquad \text{as} \: x \to b^-,
\end{equation}
 the limiting distribution of $\lambda_1$ is
\begin{equation}
\lim_{\theta \to \infty} \P_{\theta\gamma}\bigg( \frac{\lambda_1 - b\theta}{(d\theta)^{\frac{1}{2m+1}}} < s\bigg) = F_{2m+1}^2(s).
\end{equation}

From the Schur measure, we can also prove the exact Borodin--Okounkov expression~\eqref{eq:boformula} for the distribution $\P_{\theta\gamma}(k_{\max} <\ell)$. For each positive integer $\ell$, we have
\begin{equation} \label{eq:sumoverschur}
 \P_{\theta\gamma}(k_{\max} <\ell) = \P_{\theta\gamma}(\lambda_1 \leq \ell) = e^{-\theta^2 \sum_{r} r \gamma_r^2}\sum_{\lambda: \lambda_1 \leq \ell } s_\lambda [\theta\gamma]^2.
 \end{equation} 
The Schur function $s_\lambda[t]$ can be written as the $\lambda_1 \times \lambda_1$ determinant
\begin{equation}
s_\lambda[t] = \det_{1 \leq i,j \leq \lambda_1} e_{\lambda_i - i + j}[t] , \qquad \text{where } \: \sum_{n \in \Z} e_{n}[t]z^n = e^{-\sum_r (-1)^rt_r z}
\end{equation}
(this is the dual Jacobi--Trudi identity, and the  $e_n$ are the elementary symmetric functions in the Miwa times $t_r$). Since we have $e_{0} = 1$ and $e_n = 0$ for $n < 0$, we see that $s_\lambda$ can equivalently be written as $\det_{1 \leq i,j \leq \ell} e_{\lambda_i - i + j}$ for any $\ell \geq \lambda_1$. Then we see that the final expression in~\eqref{eq:sumoverschur} is proportional to a sum over $\ell \times \ell $ minors, and we have
\begin{equation} \label{eq:toeplitzz}
\sum_{\lambda: \lambda_1 \leq \ell } \det_{1 \leq i,j \leq \ell} e_{\lambda_i - i + j}[\theta\gamma] \det_{1 \leq i,j \leq \ell} e_{\lambda_i - i + j}[\theta\gamma] = \det_{1 \leq i,j \leq \ell} \sum_{n \in \Z}e_{n - i + j}[\theta\gamma] e_{n} [\theta\gamma]
\end{equation}
by the Cauchy--Binet identity. This is a Toeplitz determinant, and the generating function of the matrix entries is
\begin{equation}
\sum_{m \in \Z} z^m \sum_{n \in \Z}e_{n + m}[\theta\gamma] e_{n} [\theta\gamma] = \sum_{m,n \in \Z} z^{n+m} e_{n + m}[\theta\gamma] z^{-n} e_{n} [\theta\gamma] = e^{-\theta \sum_r (-1)^r \gamma_r (z^r + z^{-r })}. 
\end{equation}
Inserting the Toeplitz determinant~\eqref{eq:toeplitzz} into~\eqref{eq:sumoverschur} gives us precisely the formula~\eqref{eq:boformula}.

\bibliography{bib_splitfermi.bib}

\end{document}